\documentclass[iop]{emulateapj}

\newcommand{\be}{\begin{enumerate}}
\newcommand{\ee}{\end{enumerate}}

\newcommand{\lyman}{Lyman-$\alpha$}
\newcommand{\hi}{\ion{H}{1}}
\newcommand{\hii}{\ion{H}{2}}
\newcommand{\cii}{\ion{C}{2}}
\newcommand{\ciii}{\ion{C}{3}}
\newcommand{\civ}{\ion{C}{4}}
\newcommand{\feii}{\ion{Fe}{2}}
\newcommand{\nii}{\ion{N}{2}}
\newcommand{\niii}{\ion{N}{3}}
\newcommand{\sii}{\ion{Si}{1}}
\newcommand{\siii}{\ion{Si}{2}}
\newcommand{\siiii}{\ion{Si}{3}}
\newcommand{\siiv}{\ion{Si}{4}}
\newcommand{\mgi}{\ion{Mg}{1}}
\newcommand{\mgii}{\ion{Mg}{2}}
\newcommand{\mgx}{\ion{Mg}{10}}
\newcommand{\ovi}{\ion{O}{6}}
\newcommand{\ovii}{\ion{O}{7}}
\newcommand{\neviii}{\ion{Ne}{8}}
\newcommand{\msun}{\ifmmode {\rm M}_{\odot} \else M$_{\odot}$\fi}
\newcommand{\lsun}{\ifmmode {\rm L}_{\odot} \else L$_{\odot}$\fi}
\newcommand{\kms}{\ifmmode \mbox{\,km\,s}^{-1} \else \,km\,s$^{-1}$\fi}
\newcommand{\msfr}{\dot{M}_{\rm{sfr}}}
\newcommand{\mstar}{M_{\star}}
\newcommand{\mstarnot}{M_{\star,0}}

\newcommand{\mdust}{M_{\rm dust}}

\newcommand{\rperp}{R_{\perp}}
\newcommand{\rvir}{R_{\rm vir}}

\newcommand{\logoh}{$\log(\mbox{O}/\mbox{H})$}
\newcommand{\tlogoh}{$12+\log(\mbox{O}/\mbox{H})$}

\newcommand{\zg}{Z_{\rm g}}
\newcommand{\zstar}{Z_{\star}}
\newcommand{\zsun}{Z_{\odot}}

\newcommand{\alphafe}{[\alpha/\mbox{Fe}]}

\newcommand{\mzism}{M_{Z,{\rm ism}}}
\newcommand{\moxyism}{M_{{\rm oxy,ism}}}
\newcommand{\moxydust}{M_{{\rm oxy,dust}}}

\newcommand{\mstarz}{M_{Z,\star}}
\newcommand{\mstaroxy}{M_{\rm oxy,\star}}
\newcommand{\foxyretain}{f_{\rm oxy,retain}}
\newcommand{\fzretain}{f_{\rm Z,retain}}
\newcommand{\mg}{M_{\rm g}}

\newcommand{\fg}{F_{\rm g}}

\newcommand{\etal}{et~al.}
\newcommand{\oned}{1-{\sc D}}

\newcommand{\mzr}{mass-metallicity relation}

\newcommand{\rmxaa}{RMxAA}

\usepackage{verbatim}
\usepackage[usenames]{color}
\usepackage[bookmarks=false, colorlinks=true, citecolor=blue, backref=true]{hyperref}
\definecolor{dkgreen}{RGB}{0,200,0}

\shorttitle{Metal Budgets and Accounting}
\shortauthors{Peeples et al.}
\begin{document}

\title{A Budget and Accounting of Metals at $z\sim 0$: Results
    from the COS-Halos Survey\altaffilmark{1}}

\author{
  Molly S.\ Peeples\altaffilmark{2,3},
  Jessica K.\ Werk\altaffilmark{4}, 
  Jason Tumlinson\altaffilmark{3}, \\
  Benjamin D.\ Oppenheimer\altaffilmark{5,6}, 
  J. Xavier Prochaska\altaffilmark{4}, 
  Neal Katz\altaffilmark{7},
  David H.\ Weinberg\altaffilmark{8}
}
\email{molly@stsci.edu}

\altaffiltext{1}{Based on observations made with the NASA/ESA Hubble
  Space Telescope, obtained at the Space Telescope Science Institute,
  which is operated by the Association of Universities for Research in
  Astronomy, Inc., under NASA contract NAS 5-26555. These observations
  are associated with program GO11598.}
\altaffiltext{2}{Southern California Center for Galaxy Evolution Fellow, Department of Physics and Astronomy, University of California, Los Angeles, Los Angeles, CA}
\altaffiltext{3}{Space Telescope Science Institute, Baltimore, MD}
\altaffiltext{4}{UCO/Lick Observatory, University of California, Santa Cruz, CA}
\altaffiltext{5}{CASA, Department of Astrophysical and Planetary Sciences, University of Colorado, Boulder, CO}
\altaffiltext{6}{Leiden Observatory, Leiden University, the Netherlands}
\altaffiltext{7}{Department of Astronomy, University of Massachusetts, Amherst, MA}
\altaffiltext{8}{Department of Astronomy, The Ohio State University, Columbus, OH}

\begin{abstract}
  We present a budget and accounting of metals in and around
  star-forming galaxies at $z\sim 0$.  We combine empirically derived star
  formation histories with updated supernova and AGB yields and rates
  to estimate the total mass of metals produced by galaxies with
  present-day stellar mass of $10^{9.3}$--$10^{11.6}\,\msun$.  On the
  accounting side of the ledger, we show that a surprisingly constant 20--25\% mass
  fraction of produced metals remain in galaxies' stars, interstellar
  gas and interstellar dust, with little dependence of this fraction
  on the galaxy stellar mass  (omitting those metals immediately locked
  up in remnants). Thus, the bulk of metals are outside of
  galaxies, produced in the progenitors of today's $L^*$ galaxies.
  The COS-Halos survey is uniquely able to measure the mass of metals
  in the circumgalactic medium (to impact parameters of $<
  150$\,kpc) of low-redshift $\sim L^*$ galaxies. Using these data, we
  map the distribution of CGM metals as traced by both the highly
  ionized \ovi\ ion and a suite of low-ionization species; combined
  with constraints on circumgalactic dust and hotter X-ray emitting
  gas out to similar impact parameters, we show that $\sim 40$\% of
  metals produced by $\mstar\sim 10^{10}\msun$ galaxies can be
  easily accounted for out to 150\,kpc. With the current data, we
  cannot rule out a constant mass of metals within this fixed physical
  radius.  This census provides a crucial boundary condition for the
  eventual fate of metals in galaxy evolution models.
\end{abstract}


\section{Introduction}\label{sec:intro}
As all elements heavier than boron are produced by stars and
supernovae, the eventual fate of heavy elements (``metals'') is a
unique boundary condition for galaxy evolution models.  Stellar winds
and supernovae expel metals into the interstellar medium (ISM) or from
galaxies via large-scale outflows.  The metal-enriched ISM continues
to cool, collapse, and form new stars, thereby trapping some metals in
stars.  The relative distribution of metals in stars, in the ISM, and
outside of galaxies is highly sensitive to galaxies' star formation
histories, outflow histories, and the depths of their potential wells.
We present here an empirical budget and accounting of metals both
inside of star-forming galaxies and in the surrounding circumgalactic
medium (CGM) at $z\sim 0$ as a function of galaxy mass.

Early attempts to take census of metals at higher redshifts
\citep{pettini99,ferrara05,bouche05,bouche06,bouche07}
found that the bulk of metals produced by these galaxies have been
expelled by $z\sim 2$. 
Contemporary studies of $0.5\lesssim z\lesssim
5$ damped \lyman\ systems, despite their being a biased tracer of
cosmic metals, likewise inferred metal mass densities a
factor of ten lower than that expected from the star
formation history of the universe \citep{prochaska03}.  
Given the near-ubiquity of galaxy-scale
outflows at these redshifts \citep{shapley03,weiner09}, it follows that
the majority of metals produced in galaxies may have been transported
away from their stellar components (see also \citealp{lehner14}).
At lower redshifts, 
\citet{gallazzi08} found that disk-dominated, i.e., presumably
star-forming, galaxies have less than 25\% of their metals in stars.
\citet{bouche07} suggested that at $z\sim 0$ there could be as many
metals in an \ovi-traced warm-hot intergalactic medium (WHIM) as in
stars \citep[cf.][]{pagel08}. Following these ideas,
\citet{zahid12b} combined an estimate of oxygen masses in stars and
the interstellar medium at $z\sim 0$, finding that the mass of oxygen expelled from
$\log\mstar/\msun\sim 11$ galaxies is higher than the lower limit
given by the \ovi-traced CGM oxygen mass found by
\citet{tumlinson11}.
From a theoretical standpoint,
highly efficient winds that remove large amounts of freshly produced
metals are necessary to understand the ISM abundances of both
local and high-redshift galaxies
\citep[e.g.,][]{dalcanton07,erb08,finlator08,arrigoni10t,peeples11,dave11,dayal13}.
Similarly, outflows appear necessary to reach the level of observed CGM
and intergalactic medium (IGM) metal enrichment
\citep{oppenheimer06,scannapieco06,shen10,shen12,booth12,oppenheimer12,stinson12,ford13,crain13,berry13}.

Though it is possible that a significant fraction of cosmic metals are
in the CGM and IGM, it has been difficult to do a full accounting in
the low redshift Universe. Rest-frame UV spectra---and thus
space-based observations---are needed in order to probe the dominant
ionic transitions and infer column densities, mass budgets, and
kinematics.  The COS-Halos survey (GO 11598; PI J.\ Tumlinson) was
defined to address this problem with a sample of 44 galaxies observed
by HST/COS along QSO sightlines passing within 150 physical kpc of
$z\sim 0.25$ galaxies over a range of stellar mass and star formation
properties. The rationale and design for COS-Halos are described in
\citet{tumlinson13}; the key findings on metal lines are found in
\citet{tumlinson11} for high-ionization gas and in \citet{werk13} for
low-ionization gas. \citet{tumlinson11} reported that the mass of
oxygen traced by the highly ionized \ovi\ in the CGM of star-forming
galaxies is comparable to the mass of oxygen in their ISM.
COS-Halos has specifically addressed the metal
content of diffuse halo gas. The current study aims to place these
findings on highly ionized oxygen into the larger context of metal
masses contained within all galactic and circumgalactic components: stars, dust, and the
other gaseous components; we also for the first time asses the metal
mass in the lower-ionization state CGM using the COS-Halos data.

By $z=0$ the bulk of heavy elements that have ever been produced are
locked up in stellar remnants, such as white dwarfs or neutron stars.
This inventory, however, is dominated by metals that have never been
processed through the interstellar medium, and are instead directly
locked up in these remnants shortly after being produced
\citep{fukugita98,fukugita04}.  As we are interested here in constraining large-scale
gas flows and galaxy evolution processes, we ignore those metals that
are produced in stars but immediately locked into compact remnants, and consider only those heavy elements that
have been expelled into the ISM by supernovae and
stellar winds. 

Traditionally, the so-called ``missing metals problem'' has been
phrased by comparing the summed-up cosmic density of metals that
produced by the previous epochs of star formation to the cosmic
density of metals accounted for in stars the ISM, and the IGM; these
comparisons show that somewhere between 35 and 90\%\ of metals are
unaccounted for \citep[e.g.,][]{ferrara05,bouche07,pagel08}. This
large range stems from uncertainties in the cumulative star formation
rate and the cosmic density of metals in different components. Here,
we instead consider the distribution of metals produced in
star-forming galaxies as a function of stellar mass; this approach
allows us to systematically consider how uncertainties in input
scaling relations affect our census in different galaxy mass
ranges. Moreover, a mass-dependent metal inventory lends insight into
how galaxies 
with differing histories of star formation and outflow efficiency
have redistributed their metals differently through cosmic time.

We begin our census in \S\,\ref{sec:yields} with the budget of
available metals from supernovae and asymptotic giant branch (AGB)
stars.  In \S\,\ref{sec:gals} we consider the metals that are still in
galaxies, and in \S\,\ref{sec:cgm} we turn to those metals observed in
the circumgalactic medium.  We consider possible uncertainties in
these measurements as we assess the severity of the missing-metals
problem in \S\,\ref{sec:missing}, and we summarize our
conclusions in \S\,\ref{sec:conc}.  In the Appendix, we address the
implications of a non-global $\alphafe$ ratio with a similar
accounting for oxygen.

Throughout we assume a \citet{chabrier03b} initial mass function
(IMF).
Our choice of IMF mainly affects our inferred rates of metal
production via, for example, the supernova rate relative to the star
formation rate.  We also assume that all galaxies have---and have
always had---this same IMF, though there is increasing evidence in the
literature that the true picture may be more complicated
\citep[e.g.,][]{auger10,brewer12,conroy12,sonnenfeld12,dutton13,geha13}. 
Where relevant, we adopt Solar abundances from \citet{caffau11} and \citet{lodders09}.

\section{The budget of available metals}\label{sec:yields}
We first turn to the $z\sim 0$ metal budget, i.e., how many metals
have galaxies made available?  We consider the entire inventory of
metals ejected from stars through processes including explosive
supernovae and stellar winds.  We ignore metals that remain locked in
stellar remnants including white dwarfs, neutron stars, and black
holes. Though these sources comprise the majority of metals ever
produced over cosmic time \citep{fukugita04}, doing a precise and
accurate budget and accounting for remnants is 
difficult, and is insensitive to the questions of galaxy gas
flows that we wish to address here.\footnote{Throughout this paper,
  when we refer to the mass of metals ``made'', ``produced'', or ``available'', we
  are implicitly ignoring the metals directly locked up in remnants and
  instead only considering those metals expelled by supernovae and AGB
  stars.}

\begin{figure}
\includegraphics[width=0.48\textwidth]{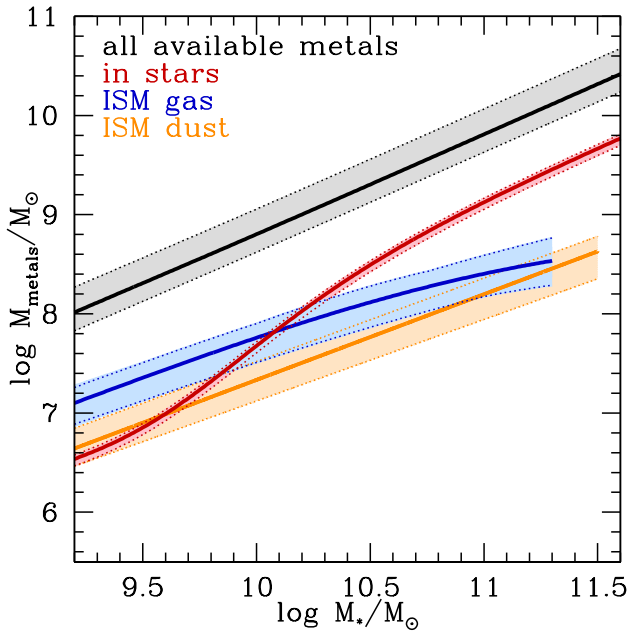}
\caption{\label{fig:mz}The total mass of metals produced by supernovae
  and AGB stars (\S\,\ref{sec:yields}; black), the mass of metals
  currently in stars (\S\,\ref{sec:stars}; red), the mass of metals in
  interstellar gas (\S\,\ref{sec:ism}; blue), and the mass of metals
  in dust (\S\,\ref{sec:dust}; orange), all plotted vs.\ the $z=0$
  galactic stellar mass of star-forming galaxies.  The shaded regions and dotted lines denote
  the adopted uncertainties as discussed in the text.}
\end{figure}

By mass, roughly 80--85\% of the metals made come from core-collapse
supernovae (\S~\ref{sec:snii}), with the rest being made by Type~Ia
supernovae (\S~\ref{sec:snia}) and AGB stars (\S~\ref{sec:agb}). In
contrast, essentially all of the oxygen comes from core-collapse
supernovae, with the small contribution made by Type~Ia supernovae
subsequently destroyed by AGB stars in massive galaxies.  The black
line in Figure~\ref{fig:mz} shows all of the metals produced, i.e.,
$M_{\rm z,ii} + M_{\rm z,ia} + M_{\rm z,agb}$, with the shaded grey
region denoting our adopted uncertainties, as described below. (The
colored regions denote metals in galaxies, as described in \S\,\ref{sec:gals}.)

The mass of metals produced by a galaxy depends on its historical rates
of Type~II supernovae, Type~Ia supernovae, and AGB stars. While these
rates can be estimated from the $z=0$ stellar mass, each is
explicitly tied to the galaxy's star formation history.  Therefore, we
use star formation histories derived by \citet{leitner12} to model the
rates of metal production.\footnote{We note that \citet{leitner12}
  adopts a slightly different cosmology than used for the COS-Halos
  sample in \S\,\ref{sec:cgm}, but that this should not have a
  substantial effect on our results.}  By assuming that star-forming galaxies
have evolved along the mean observed star forming sequence (i.e., the
evolving $\msfr$-$\mstar$ relation), \citeauthor{leitner12}\ derived
star formation rates as a function of redshift for $z=0$ galaxies.
This straightforward empirical model works well to reproduce the star
formation histories inferred from the fossil record of $\sim L^*$
galaxies, although it does predict too {\em much} downsizing for
$\mstarnot\lesssim 10^9\,\msun$ galaxies (i.e., it predicts that these
dwarf galaxies have formed all of their stars at late times, contrary
to what is observed).  Fortuitously, the bulk of
metals are produced through core-collapse supernovae, and
the final mass of metals produced by Type~II supernovae depend more sensitively on the total
mass of stars formed than on the precise history of when the stars
formed (unlike the products of Type~Ia supernovae or AGB stars).  Hence, our results are not
very dependent on the assumed star formation histories, although these
histories do provide a useful framework for examining the sources of
metals in galaxies of different mass.

\subsection{Core-collapse Supernovae}\label{sec:snii}
The mass of metals made by Type~II supernovae (i.e.,
core-collapse supernovae) is
\begin{equation}
M_{\rm Z,ii} = \int y_{{\rm z,ii}}\,\msfr\,{\rm d}t
\end{equation}
where $\msfr$ is the rate at which stars of mass 0.1--100\,\msun\ are
being made, and $y_{{\rm z,ii}}$ is the
nucleosynthetic yield of all heavy elements
produced by Type~II supernovae.  Because not all stars made survive to
$z=0$ (i.e., $\mstarnot<\int \msfr\,{\rm d}t$), the mass of metals
made by Type~II supernovae is higher than $y_{{\rm z,ii}}\mstarnot$,
where $\mstarnot$ is the galaxy mass at $z=0$.  Specifically, using
the star formation histories provided by \citet{leitner12}, we find
that the Type~II supernova metal production can be well described as
\begin{equation}
\log (M_{\rm z,ii}/\msun) = 1.0146\log(\mstarnot/\msun) + \log y_{{\rm z,ii}} + 0.1091.
\end{equation}
  Note that this relation is not very
dependent on the assumed star formation history; that is, the slope is
nearly unity. For the mass range we
consider here, this is quite close to just assuming that all galaxies
have recycled $\sim55$\% of their stellar mass (i.e., $\int
\msfr\,{\rm d}t\sim 1.8\times\mstarnot$),\footnote{More specifically,
  $ 1.0146\times 10 + \log
  y_{{\rm z,ii}} + 0.1091 \approx\ 10  + \log 1.8+ \log
  y_{{\rm z,ii}}$.} which for a
\citet{chabrier03b} IMF is very similar to the commonly used
``instantaneous recycling approximation'' for all stars $m>1\msun$.

We adopt nucleosynthetic yields of
$y_{{\rm z,ii}}=0.030$, with the main elemental contributions having
yields of 0.015 (oxygen), 0.0083 (carbon),
0.0014 (silicon), 0.0011 (iron), and 0.001 (nitrogen). These values are in the middle of the range
we derive from the non-primordial models of
 \citet{woosley95}, \citet{portinari98}, \citet{chieffi04}, and \citet{hirschi05},
under the assumption that stars of mass $10<m<100\,\msun$ end as
core-collapse supernovae.  Our adopted yields are close to those given
by \citet{chieffi04}.  The grey shaded region in Figure~\ref{fig:mz}
includes the effects of letting our assumed yields vary within the
ranges given by the models: we let $y_{{\rm z,ii}}$ vary from $0.0214$
to $0.0408$, and let $y_{{\rm o,ii}}$ vary from $0.01394$ to $0.01828$
(see the Appendix).  \citeauthor{woosley95} tend to give lower
$y_{{\rm z,ii}}$ than the more recent models; we note that this means
our fiducial yields are somewhat higher than those assumed in earlier
chemical evolution models \citep[e.g.,][]{madau96b,ferrara05}.
On the other hand, the only model that
includes stellar rotation \citep{hirschi05} gives a much higher metal yield
than those neglecting rotation.
As the models give little dependence of the yield on the progenitor
metallicity---especially given the full range of uncertainty in the
above models---we do not take into account a galaxy's metallicity
history and instead assume that the yields do not vary.

For a single population of stars with mass $0.1<m<100\,\msun$ formed
with a \citet{chabrier03b} IMF, 18.58\%\ of the mass is in stars with
$10<m<100\,\msun$ and 21.16\%\ in stars with
$8<m<100\,\msun$.\footnote{These fractions will be lower by a factor
  of $\sim 3$ for a
  \citet{salpeter55} IMF.}  The uncertainties shown in
Figure~\ref{fig:mz} include the effects of letting the minimum
supernova mass vary from $8\,\msun$ to $10\,\msun$.

\subsection{Type~Ia Supernovae}\label{sec:snia}
We adopt a $t^{-1}$ delay time distribution for Type~Ia supernovae,
where
\begin{equation}\label{eqn:Iarate}
{\rm SNR}(t)_{\rm Ia} = 4\times 10^{-13}\,{\rm yr}^{-1}\,m_{\star}\left(\frac{t}{1\,{\rm Gyr}}\right)^{-1}
\end{equation}
is the rate of Type~Ia supernovae produced by a stellar population of
mass $m_{\star}$ formed at time $t=0$ \citep{maoz12a}.  The metal
production from Type~Ia supernovae is then described by
\begin{equation}
M_{\rm Z,ia} = \int m_{{\rm z,ia}}\,{\rm SNR}_{\rm Ia}\,{\rm d}t,
\end{equation}
where $m_{{\rm z,ia}}=1.2256\,\msun$ 
is the mass of metals produced per Type~Ia supernova; in the Appendix,
we assume that each Type~Ia supernova produces $0.143\,\msun$ of oxygen
\citep{thielemann86,tsujimoto95}.  Fifty percent of the
total metal mass is in the form of iron.  The mass of metals 
oxygen produced by Type~Ia supernovae are well described as
\begin{eqnarray}
\log (M_{\rm z,ia}/\msun) &=& 1.043\log(\mstarnot/\msun) - 2.683.
\end{eqnarray}

Though the Type Ia metal production is more dependent on the assumed
star formation histories than it is for core-collapse supernovae
(\S\,\ref{sec:snii}), because Type~Ia supernovae are responsible for
only $\sim$10\% of the metals, uncertainties in the star formation
histories or Type~Ia supernova rates do not strongly affect our
integrated results.  The grey shaded region in Figure~\ref{fig:mz}
takes into account a factor of two uncertainty in the Type~Ia
supernova rate \citep{maoz12a} and lets the masses of metals produced
vary by a factor of two \citep{gibson97}.

\subsection{Asymptotic Giant Branch Stars}\label{sec:agb}
Metal processing by AGB stars depends sensitively on their mass and
initial metallicity \citep[e.g.,][]{karakas10}.  Following
\citet{peeples13}, we determine a galaxy's evolving metallicity by
assuming that as galaxies form stars according to the
\citet{leitner12} star formation histories, the ISM metallicity
remains on the observed $z\sim 0$ relation between stellar mass, star
formation rate, and gas-phase metallicity as measured by
\citet{mannucci10}.  This assumption is motivated by the fact that
observed galaxies at $0\lesssim z\lesssim 2.2$ all appear to lie on
the same relation.  \citet{peeples13} showed that this simple approach
for modeling galaxy metallicity histories reproduces the $z=0$ stellar
metallicities as measured by \citet{gallazzi05} and \citet{woo08}
rather well.

We adopt the AGB metal yield look-up tables as tabulated by
\citet{oppenheimer08} for their Gadget-2 simulations and apply those
yields to our model star formation histories, assuming a factor of two
uncertainty in the yields.  They tabulated AGB
yields from stellar models of different input masses and metallicities
calculated by \citet{marigo01}, \citet{herwig04}, and
\citet{gavilan05}, and converted from stellar mass to age using
stellar lifetimes, since the AGB phase occurs right before stellar
death.  We use these tables as a function of age and metallicity to
calculate the amount of carbon, nitrogen, and oxygen produced by stars
dying at each $1.3\times10^7$ yr time step using the \citet{leitner12}
star formation histories and the \citet{peeples13} stellar
metallicities.  We are interested in the total amount of C, N, and O
``processed'' by the AGB stars, which is the ejected metal mass in
excess of the metals present in the star at birth.  Significant
amounts of carbon and nitrogen are produced in AGB stars compared to
Type II SNe production.  Oxygen can be lost via AGB stars because this
element is burned, remaining trapped in the remnant, in many AGB stars
\citep{karakas12}. However, the net loss is small compared to the
amount of oxygen produced by Type II SNe. Over the mass range and for
the AGB models we
consider here, oxygen is primarily destroyed in AGB stars, and the
total change in metal mass owing to AGB processing is mainly from
added carbon mass.  Other metals are not significantly processed by
AGB stars when considering the mass of all metals: metals heavier than
oxygen mostly remain unprocessed, while the species with the largest
changes are rare isotopes by mass (e.g. $^{13}$C).

\subsection{Summary: Metal Budgets}\label{sec:budget}
\begin{figure}
\includegraphics[width=0.48\textwidth]{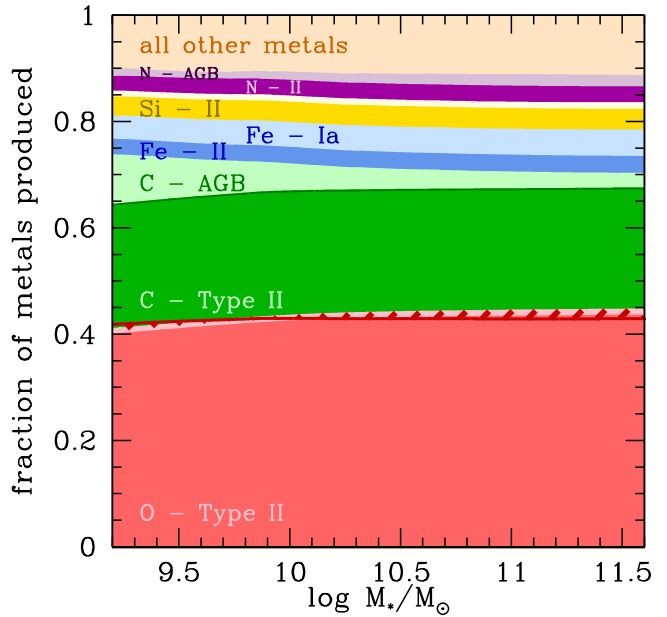}
\caption{\label{fig:elements} The cumulative fraction of metals
  produced by $z=0$, broken down by element and source vs.\ the galaxy
  stellar mass.  As most metals are produced in Type~II supernovae,
  the variations in elemental fraction with respect to stellar mass
  due to varying star formation histories in contributions from
  Type~Ia SNe and AGB stars are relatively small.  The most abundant
  element is oxygen (dark pink, Type~II SNe; light pink, Type~Ia SNe;
  dark red hashed, AGB, showing the destruction of oxygen in more
  massive galaxies), followed by carbon (medium green, Type II SNe;
  dark green, Type Ia SNe; pale green, AGB), iron (dark blue, Type II
  SNe; pale blue, Type Ia SNe), silicon (dark yellow, Type II SNe;
  pale yellow, Type Ia SNe), and nitrogen (dark purple, Type II SNe;
  pale purple, AGB).  The contribution from other elements is $\sim
  10$\% (pale orange).  }
\end{figure}

Figure~\ref{fig:elements} shows the breakdown of metal production
sources by element (O, C, Fe, Si, and N) as a function of $\mstarnot$.
Interestingly, the expected fraction in oxygen is very close to the Solar
value of $\sim 44$\%; in fact, to within a few percent, the pattern
shown in Figure~\ref{fig:elements} is close to the Solar abundance
pattern, with the obvious exception of Ccarbon (the Solar C/Z is
$\approx 18$\% as opposed to the $\sim 30$\% shown in
Fig.\,\ref{fig:elements}). We note, however, that 
in addition to the AGB yields being highly uncertain,
the channel for
carbon production from core-collapse supernovae is poorly understood;
among the different supernova models we considered, the carbon yield
varies by over a factor of 4 (as opposed to the factor of $< 2$ for
all metals).

\section{How many metals are still in galaxies?}\label{sec:gals}
We now turn to an accounting of metals at $z\sim 0$ by considering
those metals that are in galaxies.  The metals in galaxies are shown
in Figure~\ref{fig:mz}: we consider metals in stars (red,
\S\,\ref{sec:stars}), the neutral interstellar medium (ISM; blue,
\S\,\ref{sec:ism}), or interstellar dust (orange,
\S\,\ref{sec:dust}).  As elsewhere in this paper, we assume here that
we can use mean scaling relations to describe a ``typical'' galaxy and
that the scatter about these scaling relations does not affect our
conclusions about these typical galaxies.  For the largest two
contributors, stars and the ISM, the relevant observations are from
the Sloan Digital Sky Survey (SDSS); thus we further assume that the
3\arcsec\ subtended by SDSS fibers are representative of the global
stellar or gaseous metallicities (and that accurate aperture
corrections have been made).

\subsection{Stars}\label{sec:stars}
The mass of metals in stars at $z=0$ of a galaxy with stellar mass $\mstarnot$ is
\begin{equation}\label{eqn:stars}
\mstarz = \zstar\mstarnot.
\end{equation}
The observed stellar metallicity $\zstar$ is weighted by the most luminous stars in the
galaxy, which are also the youngest, and, therefore, most metal rich stars.
\citet{peeples13} traced the buildup of metals in stars by following
the \citet{leitner12} star formation histories (see \S\,\ref{sec:agb}).
Using the stellar population synthesis models of \citet{bruzual03}
they compared the $B$-band weighted stellar metallicities,
$Z_{\star,\rm light}$, to the more physical mass-weighted stellar
metallicities, $Z_{\star,\rm mass}$.  This correction is well fit by
the power law
\begin{equation}\label{eqn:mtl}
\log (Z_{\star,\rm mass}/\zsun) = 1.08\log (Z_{\star,\rm light}/\zsun) - 0.16.
\end{equation}
We correct the observed stellar metallicities to a mass-weighted value
using equation~(\ref{eqn:mtl}).

The red line in Figure~\ref{fig:mz} denotes the mass of metals
locked up in stars, as a function of $\mstar$, using the $Z_{\star}$
measurements from \citet{gallazzi05}; uncertainties in these
descriptions of ``typical'' $\zstar$ are not shown.
We use equation~(\ref{eqn:mtl}) to convert from luminosity-weighted
observations to the actual masses.  The band shows the uncertainty in
the solar abundance.  We adopt $\zsun\equiv M_{Z,\odot}/\msun=0.0153$
from \citet{caffau11} and show the range of $\zsun=0.013$ to $0.0168$
from \citet{bahcall05} as possible sources of uncertainty. Compared to
the uncertainties relevant to other components, the contribution from
the uncertainty in the Solar abundance is small.  \citet{gallazzi05}
give an uncertainty of $\sim0.12$\,dex in metallicity for each galaxy,
though their stated medians have much smaller uncertainties.

\subsection{Interstellar Gas}\label{sec:ism}
The mass of metals in the interstellar medium at $z=0$ is simply
\begin{equation}\label{eqn:ism}
\mzism = \zg\mg,
\end{equation}
where $\zg$ is the metallicity of the ISM and $\mg$ is the galaxy gas
mass.  Generally the radial metallicity gradients in spiral galaxies
at large radius are shallow, implying that the gas is well-mixed
\citep{cartledge04,werk11}. Hence, for the gas masses, we include all of
the cold gas, i.e., everything that is traced by \hi\ plus molecular gas.

\begin{figure}
\includegraphics[width=0.48\textwidth]{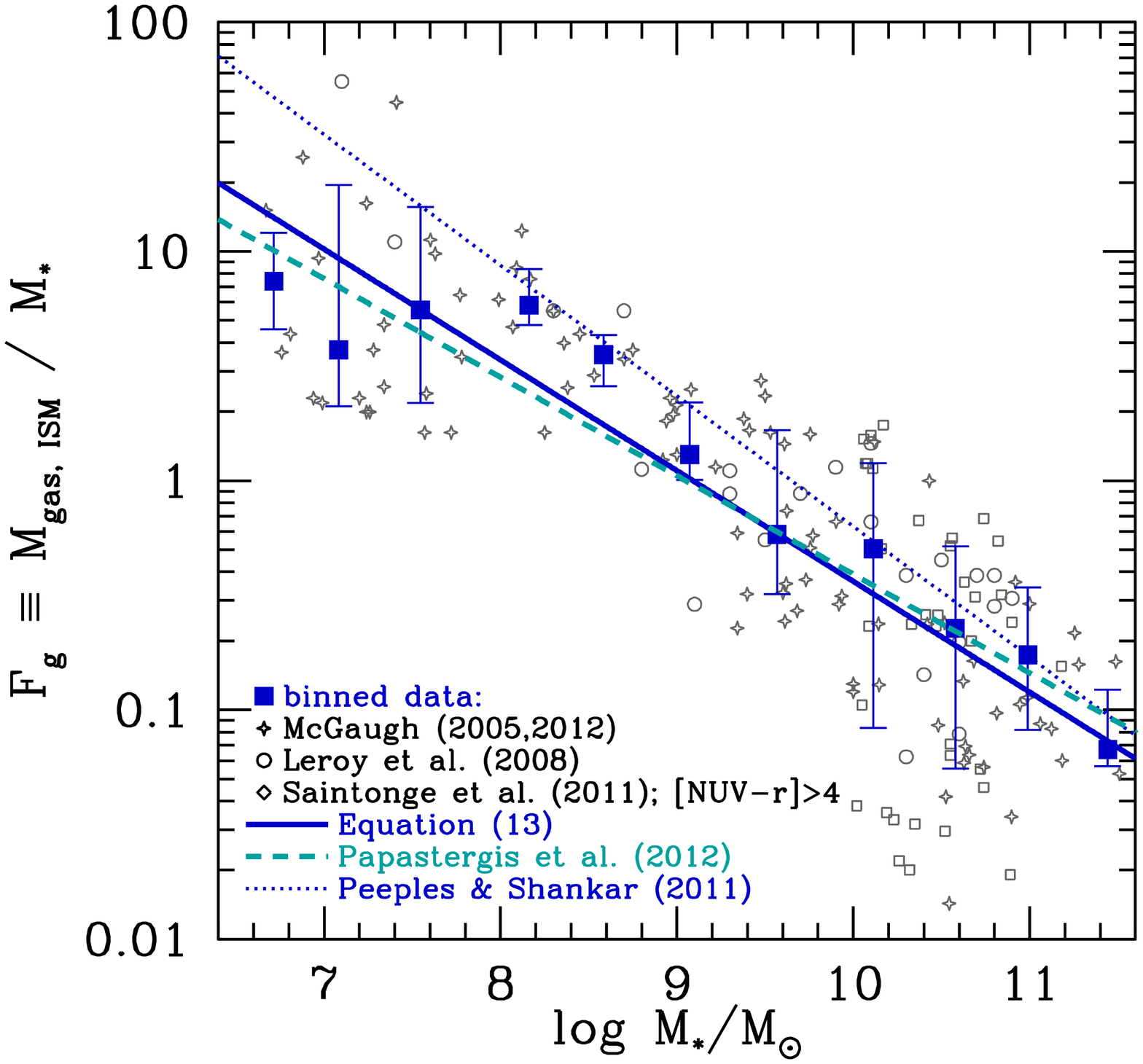}
\caption{\label{fig:Fgas}Cold (atomic $+$ molecular) gas fractions,
  $\fg\equiv\mg/\mstar$, as a function of stellar mass.  The grey data points
  are collated from \citet{mcgaugh05,mcgaugh12}, \citet{leroy08}, and
  \citet{saintonge11}, with selections for star-forming galaxies and a
  conversion to a \citet{chabrier03a} IMF where appropriate. The blue
  points are the median $\fg$ in bins of $\Delta\log\mstar=0.5$\,dex,
  with errorbars denoting the 16- to 84\%\ range of the data, as
  tabulated in Table~\ref{tbl:Fgbins}.  The solid
  blue line is a fit to the data (equation~\ref{eqn:Fgas}); the dashed
  cyan line is an independent fit to similar data from \citet[][see
  also equation~\ref{eqn:papas}]{papastergis12}.  For comparison we
  also show the ``total'' gas fractions from \citet{peeples11} (dotted
  blue line), which strongly over-estimates the gas fractions for
  dwarf galaxies.}
\end{figure}

In Figure~\ref{fig:Fgas} we show the atomic plus molecular gas mass
measurements from \citet{mcgaugh05,mcgaugh12} and \citet{leroy08},
adding star-forming (${\rm NUV}-r<4$) galaxies from the COLDGASS
sample \citep{saintonge11}. We list the median, 16- and 84\%\ ranges of these
data in bins of stellar mass in
Table~\ref{tbl:Fgbins}. A least-squares fit to the full data set gives us the gas fractions
as a function of stellar mass, 
\begin{equation}\label{eqn:Fgas}
\log\fg \equiv \log (\mg/\mstar) = -0.48\log\mstar + 4.39.
\end{equation}
Using a similar approach, \citet{papastergis12} find 
\begin{equation}\label{eqn:papas}
\log\fg = \log (1.366M_{\rm HI}/\mstar) = -0.43\log\mstar + 3.89,
\end{equation}
by fitting a power law to data compiled from optically-selected galaxy
samples \citep{swaters02, garnett02, noordermeer05, zhang09}. We
convert their $M_{\rm HI}$ to the total gas mass by setting $\mg=1.366M_{\rm HI}$ to
correct for He, by adopting the Solar value for the mean atomic
weight of $1.366$ \citep{caffau11}.  We use the average of the
gas masses given by these two fits and the uncertainty ranges in
Figure~\ref{fig:mz} take into account the (small) range these fits
span at higher $\mstarnot$.

We note that \citet{peeples11} gave gas fractions that were
biased high (dotted blue line in Figure~\ref{fig:Fgas}) because many
galaxies in their sample were \hi-selected \citep{west09,west10},
which led to an overestimate of $\fg$ at fixed $\mstar$
\citep{papastergis12}.  Moreover, their ``total'' fit was also offset
by $+0.2$\,dex (for illustrative purposes) and was steeper than what we find here. The fits given
by equations~(\ref{eqn:Fgas}) and (\ref{eqn:papas}) are also lower
than those derived from the \citeauthor{peeples11} binned data, as
done by, e.g, \citet{zahid12b}.  We provide an update to these binned
data in Table~\ref{tbl:Fgbins}. As the mass of metals in the ISM is
proportional to the gas mass (equation~\ref{eqn:ism}), these seemingly
small differences can have a large impact on the missing-metals
problem, especially for gas-rich dwarf galaxies.

\begin{table}
\centering
\begin{tabular}{rrrr}\hline\hline 
$\langle{\log\mstar}\rangle$ & median $\fg$ & 16\%ile $\fg$ & 84\%ile $\fg$\\\hline
$6.7$ & $7.4\phantom{0}\phantom{0}$ & $12.0\phantom{0}$ & $4.6\phantom{0}\phantom{0}$\\ 
$7.1$ & $3.7\phantom{0}\phantom{0}$ & $19.5\phantom{0}$ & $2.1\phantom{0}\phantom{0}$\\ 
$7.6$ & $5.6\phantom{0}\phantom{0}$ & $15.7\phantom{0}$ & $2.2\phantom{0}\phantom{0}$\\ 
$8.2$ & $5.8\phantom{0}\phantom{0}$ & $8.4\phantom{0}$ & $4.8\phantom{0}\phantom{0}$\\ 
$8.6$ & $3.6\phantom{0}\phantom{0}$ & $4.3\phantom{0}$ & $2.6\phantom{0}\phantom{0}$\\ 
$9.1$ & $1.3\phantom{0}\phantom{0}$ & $2.2\phantom{0}$ & $1.0\phantom{0}\phantom{0}$\\ 
$9.6$ & $0.6\phantom{0}\phantom{0}$ & $1.7\phantom{0}$ & $0.32\phantom{0}$\\ 
$10.1$ & $0.50\phantom{0}$ & $1.2\phantom{0}$ & $0.083$\\ 
$10.6$ & $0.23\phantom{0}$ & $0.52$ & $0.056$\\ 
$11.0$ & $0.17\phantom{0}$ & $0.34$ & $0.082$\\ 
$11.4$ & $0.067$ & $0.12$ & $0.057$\\ \hline 
\end{tabular}
\caption{Median, 16-, and 84-percentile cold gas fractions $\fg\equiv\mg/\mstar$ in bins of
$\Delta\log\mstar=0.5$\,dex for the \citet{mcgaugh05,mcgaugh12}, \citet{leroy08}, and
  \citet{saintonge11} data sets, as shown in Figure~\ref{fig:Fgas}.
\label{tbl:Fgbins}}
\end{table}

In addition to the cold atomic and molecular gas traced by the data
shown in Figure~\ref{fig:Fgas}, the Milky Way has a warm ionized
medium (WIM) that may harbor a similar total and metallic mass as that
in neutral gas \citep{sembach00,haffner09}.  However, systematic
characterizations of the mass in this phase for galaxies other than
our own are rare, and it is unclear how well the WIM mass scales with
either \hi\ or stellar mass \citep{oey07}. Furthermore, the relative
amounts of ionized versus neutral gas will naturally depend on the
ambient ionization fields. We, therefore, stress that the metal masses
we quote here for ``the interstellar medium'' apply {\em only} to the
cold gas, and that the total interstellar gaseous metal masses could
be larger by as much as a factor of two.

On the other hand, because the typical $z\sim 0$ galaxy could have
a shallow metallicity gradient, or gas that is not entirely
well-mixed, the $\mg$ and thus the $\mzism$ that we adopt here should be
considered as upper-limits to the amount of metals in the cold
interstellar medium, especially at the lowest $\mstar$, where galaxies
often have very extended \hi\ envelopes relative to their stars.

\subsubsection{Oxygen in the ISM}\label{sec:oism}
The largest and most standardized surveys of ISM metallicities
relative to galaxy stellar mass are that of oxygen abundances measured from
nebular emission lines in \ion{H}{2}\ regions.  Systematic
uncertainties in theoretical photoionization models and empirical
calibrations, however, lead to dramatically different measurements for
\logoh\ for the same set of emission line fluxes; we refer the reader
to the review by \citet{kewley08} for a detailed discussion
of these issues.\footnote{These systematic uncertainties are expected
  to be resolved by adopting a $\kappa$-distribution rather than a
  Maxwell-Boltzmann distribution for electron energies
  (\citealp{nicholls12,dopita13}).}  Using SDSS
spectra, \citet{kewley08} give fits to ten mass-metallicity
relations derived from different abundance calibrations in the
literature.  We take the average of the eight most similar relations
(dropping the two with the lowest calibrations and flattest slopes,
both of which are based on $T_e$ measurements tied to the
[\ion{O}{3}]\,$\lambda\,4363$ line that are known to give
systematically low estimates of O/H).
These calibrations typically assume a $0.1$\,dex depletion of oxygen
onto dust; as we wish to isolate the {\em gas}-phase oxygen abundance,
we subtract $0.1$\,dex from each of these calibrations,
 adopting
\begin{eqnarray}\label{eqn:zg}
\log &&\frac{\moxyism}{\mg} = 27.76 - 7.056\log\mstar \\
&&+ 0.8184(\log\mstar)^2 - 0.03029(\log\mstar)^3,
\end{eqnarray}
where $\mstar$ is in units of $\msun$, and
\begin{equation}\label{eqn:tlogoh}
\frac{\moxyism}{\mg} = \frac{m_{\rm O}\times n_{\rm oxy}}{\bar{\mu}m_{\rm p}\times n_{\rm H}} = \frac{15.999}{1.366}\times\frac{n_{\rm oxy}}{n_{\rm H}},
\end{equation}
where $n_{\rm oxy}/n_{\rm H}$ is the number abundance of oxygen
relative to hydrogen as expressed by \tlogoh, $m_p$ is the mass of a
proton, $m_{\rm O}=15.999m_p$ is the mass of an oxygen atom, and we
adopt the Solar value of $\bar{\mu}=1.366$ \citep{caffau11}.  

\subsubsection{Metals in the ISM}\label{sec:zism}
Stars show a variation in $\alphafe$ relative to galactic stellar
mass (see \S\,\ref{sec:oxystar}).  Similarly, star formation driven
outflows are reasonably expected to be $\alpha$-enhanced (as they are presumably
primarily driven by Type~II supernovae, the primary sites of
$\alpha$-element formation), and the efficiency of such outflows in
removing metals from the ISM should vary as a function of galaxy mass
\citep[e.g.,][]{murray05,peeples11}.  It is, therefore, quite possible
that the interstellar $\alphafe$ varies as a function of stellar mass.
However, as definitive observations one way or the other do not exist,
we therefore adopt a Solar oxygen-to-metals ratio of
$0.00674/0.0153=0.44$ for the ISM.  The width of the blue band in
Figure~\ref{fig:mz} denotes the uncertainty in $\mzism$ and almost
entirely owes to uncertainties in the \tlogoh\ calibration 
(\S\,\ref{sec:oism}).

\subsection{Interstellar Dust}\label{sec:dust}
\begin{figure}
\includegraphics[width=0.48\textwidth]{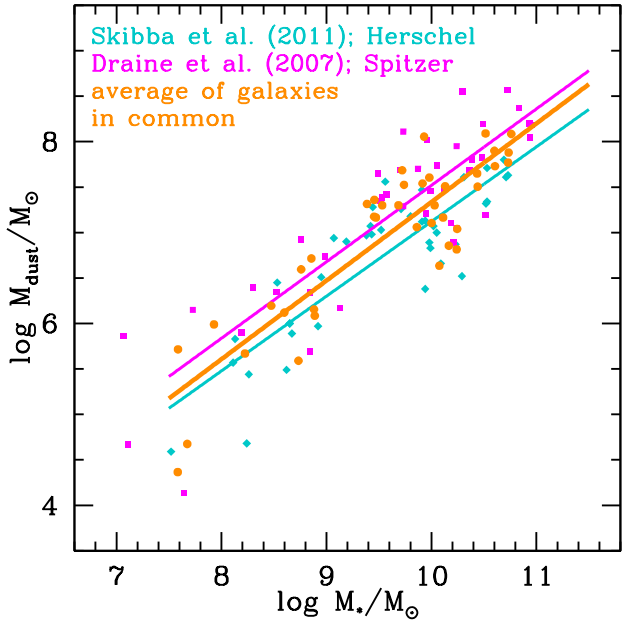}
\caption{\label{fig:dust}Dust masses versus stellar mass from a sample
  of forty star-forming galaxies, as measured by \citet{skibba11}
  ({\em cyan}) and \citet{draine07} ({\em pink}), and the average of
      the two samples ({\em orange}).}
\end{figure}

Though the absolute mass of dust in galaxies is small compared to the
total mass of stars and the ISM, dust is (by mass) almost entirely
composed of heavy elements.
Figure~\ref{fig:dust} shows measured dust masses as a function of
stellar mass from the Spitzer Infrared Nearby Galaxies Survey
\citep[SINGS;][]{draine07}, with stellar masses from
\citet{dacunha08}\footnote{We thank E.\ da Cunha for sharing these
  stellar mass measurements with us.} and from the Herschel KINGFISH
survey \citep{skibba11}.\footnote{We have corrected the
  \citet{skibba11} stellar masses by $-0.05$\,dex to convert them from
  a \citet{kroupa01} IMF to \citet{chabrier03a} IMF
  \citep{bernardi10}.}  These samples were selected to be
representative of normal galaxies at $z\sim 0$ \citep{kennicutt03b};
therefore, their dust masses should not be abnormally high (or low).
These two samples have 40 star-forming (which we define as $\msfr/\mstar >
10^{-11}$\,yr$^{-1}$) galaxies in common. The
average of the two sets of measured stellar and dust masses are shown
in orange.  We adopt the fit to the averaged data, also shown in orange,
\begin{equation}\label{eqn:dust}
\log(\mdust/\msun) = 0.86\log(\mstarnot/\msun) - 1.31.
\end{equation}
Relative to the gas fractions given in Equation~(\ref{eqn:Fgas}), this
results in a dust-to-gas ratio\footnote{The dust masses are
  measured independently of the galaxy gas masses.} of $\sim 1$\% at
$\mstarnot\sim 10^{10.5}\msun$ and a dust-to-metals ratio of $\sim
30$--50\%.  The uncertainty range shown in Figure~\ref{fig:mz}
corresponds to the two independent fits; the dotted lines are the fits
to the full data samples, and the solid lines are the independent fits
to the overlapping galaxies. We note that only 15 of the overlapping galaxies
have SCUBA sub-mm data and that these longer wavelengths are
necessary to fully constrain the mass of {\em cold} dust (see
\S\,4.2.1 of \citealt{skibba11} for more discussion).

Finally, it is possible that dust is partially destroyed in the parts
of the \hii\ regions where the emission lines used to measure
interstellar gas abundances originate. If this were the case, then by
separately counting interstellar dust and interstellar gas-phase
metals, we might possibly be double counting some of the metals.  On
the other hand, if it requires shocks from supernovae to destroy dust
grains, then the metals in dust should not be fully mixed with the gas
in \hii\ regions.
Moreover, as ISM gas abundance calibrations typically attempt to
correct for depletion of oxygen onto dust
\citep[e.g.,][]{kewley02,dopita13}, it is at least commonly assumed
that dust is not strongly destroyed in \hii\ regions.
For these reasons,
we consider it safe to take
ISM dust as a
distinct reservoir of metals, as long as the dust depletion calibrations are
taken into account when considering metals in interstellar gas.

As dust is comprised primarily of heavy elements, we simply adopt
\begin{equation}
M_{\rm Z,\,dust} = \mdust.
\end{equation}

\subsection{Summary: Metals in Galaxies}\label{sec:galsummary}
\begin{figure}
\includegraphics[width=0.48\textwidth]{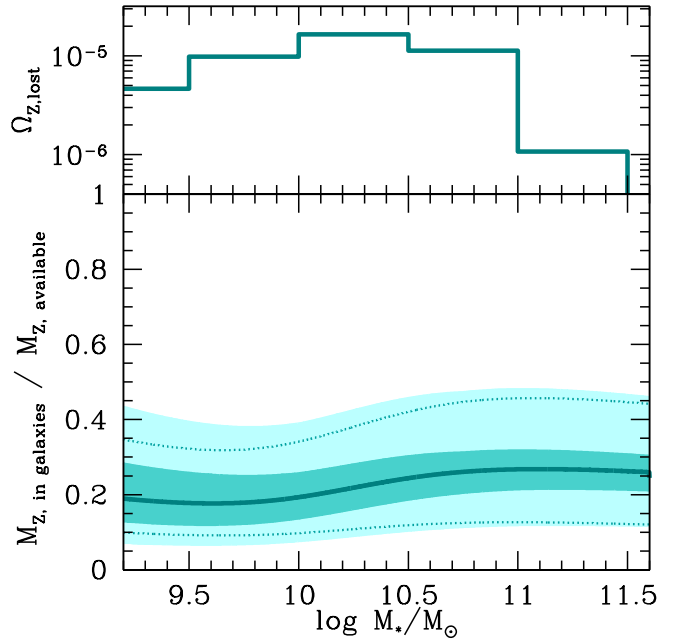}
\caption{\label{fig:flost}{\em Top}: the cosmic density of metals
  outside of star-forming galaxies, $\Omega_{\rm Z,\,lost}$, in bins of the present-day
  stellar mass of the galaxies where the metals were produced (see
  equation~(\ref{eqn:lost}); most of the expelled metals are from
  $\sim L^*$ galaxies. {\em Bottom}: the fraction of metals expelled
  by SNe and AGB stars that is retained by star-forming galaxies versus galactic
  stellar mass. The light shaded region shows the total uncertainty,
  including all sources; the inner dark shaded region, for no
  uncertainty in the budget, and the dotted lines for no uncertainty
  in the calibration of the gas-phase metallicities.}
\end{figure}

The fraction of available metals produced by star-forming $9.5
\lesssim \log\mstarnot/\msun \lesssim 11.5$ galaxies that 
they retain
is a relatively constant $\sim 20\%$. 
The bottom panel of Figure~\ref{fig:flost} shows $\fzretain\equiv M_{\rm Z,\,retain}/M_{\rm
  Z,\,made}$ versus $\log\mstarnot$ in
bins of $\log\mstarnot$,
where $M_{\rm Z,\,retain}(\mstar)$ is the mass of metals in galaxies as
stars, interstellar gas, or interstellar dust, i.e.,
\begin{equation}\label{eqn:mlost}
M_{\rm Z,\,retain}(\mstar) = \mstarz + \mzism + \mdust.
\end{equation}
The largest uncertainty in $\fzretain$ owes to
uncertainties in the metals produced, specifically from the
uncertainty in the nucleosynthetic yields ($y_{\rm z,ii}=0.030^{+0.0108}_{-0.0086}$), and, to a lesser extent,
the Type~Ia supernova rate.  
At lower masses, large gas fractions lead to an uncertainty in the
calibration of the gas-phase \mzr\ that becomes nearly as important as
the metal production uncertainty \citep[cf.][]{zahid12b}. 
With regards to ``how bad is the missing
metals problem?''  (see \S\,\ref{sec:missing}), 
the largest uncertainty still remains the mass of metals to be found.

Regardless of the normalization owing to the nucleosynthetic yields,
as $\fzretain$ is relatively constant with respect to
stellar mass, the majority of metals that have been expelled from
galaxies by $z=0$ were produced by stellar populations that are in today's $\sim
L^*$ galaxies; for star-forming galaxies, $L^*$ corresponds to $\log M_{\star}^*/\msun\sim
10.6$ \citep{ilbert13}.
The top panel of Figure~\ref{fig:flost} shows the cosmic density of metals
lost from galaxies,
\begin{equation}\label{eqn:lost}
\Omega_{\rm Z,\,lost} = \frac{1}{\rho_{\rm c}}\int\limits^{M_{\star,{\rm max}}}_{M_{\star,{\rm min}}}M_{\rm Z,\,lost}(\mstar)\;\frac{{\rm d}\,n(\mstar)}{{\rm d}\,\log\mstar}\,{\rm d}\,\log\mstar,
\end{equation}
in bins of stellar mass,
where $M_{\rm Z,\,lost}(\mstar)\equiv M_{\rm Z,\,made}(\mstar) - M_{\rm Z,\,retain}(\mstar)$.
Despite their
shallower potential wells, most of the metals outside of galaxies were
not put there by the $z\sim 0$ dwarf galaxies. However, a substantial
fraction of the metals outside of galaxies could have been produced by
dwarf galaxies at higher redshifts that were the progenitors of larger
present day galaxies \citep[see, e.g.,][]{shen12}.
Specifically, using the \citet{moustakas13}\ stellar
mass function for star-forming galaxies selected from SDSS and GALEX,
we find that 
star-forming galaxies of stellar mass $10^{8.5}$--$10^{11.5}$\,\msun\ have produced metals at a cosmic density of $\Omega_{\rm
  Z}=5.9\times 10^{-5}$, with 78\% ($\Omega_{\rm
  Z,\,lost}=4.6\times 10^{-5}$) of these metals no longer in
galaxies. Broken down by galaxy mass, $\Omega_{\rm
  Z,\,lost}$ is $2.6\times 10^{-5}$ 
for $10^{9.5}$--$10^{10.5}$\,\msun\ galaxies 
versus $7.2\times 10^{-6}$ for
$\mstar=10^{8.5}$--$10^{9.5}\,\msun$\ galaxies and $1.2\times 10^{-5}$
for $10^{10.5}$--$10^{11.5}$\,\msun\ galaxies.  
Analogous to how any given individual star is most likely to be found in an
$L^*$ galaxy, 
it is a generic result, independent of which form of the
stellar mass function is adopted,
that
the relatively constant fraction of metals expelled by galaxies
implies that
the bulk of
metals outside of galaxies at $z\sim 0$ were produced by $\sim L^*$
galaxies.

\section{Metals in The Circumgalactic Medium}\label{sec:cgm}
If most of the metals produced by galaxies can no longer be found in
those galaxies (Figure~\ref{fig:flost}), then the natural place to
seek those missing metals is in the immediate vicinity of galaxies, in
the circumgalactic medium.  As the CGM plays host to complex flows 
of gas accreting into and ejected from galaxies, 
we expect it to span a wide range of density, dynamical state, and 
ionization. We therefore consider here probes of the CGM that trace distinct
phases in temperature, density, and ionization state.   

COS-Halos addresses the component of CGM metals that can be observed
with the suite of multiphase diagnostic UV ions that it was designed
around, in the mass range of galaxies selected. As described by
\citet{tumlinson13},  COS-Halos was designed to cover galaxies both
with and without active star formation, in the range $\log
M_{\odot}/\msun \sim 9$--$11.5$.  The sample was selected to ensure coverage
of the key diagnostics of high-ionization gas and metals, and
the $\lambda\lambda 1032,1038$ doublet of \ovi\ must be observed
at $z > 0.11$ to fall on the COS FUV detectors. At the redshifts of
the COS-Halos galaxies, $0.14 <  z < 0.36$, many other diagnostic ions are also detectable, from
the neutral species \ion{O}{1}\ through low ions like \cii\ and
\siii\ and intermediate ions such as \ion{C}{3} and \ion{Si}{3}. The survey also
collected optical data with Keck/HIRES that covered the strong
$\lambda\lambda 2796,2803$ doublet of \mgii\ as well as weaker
lines of \mgi\ and \feii\ \citep{werk13}. With this wide range of
diagnostics we can address the metal budget of CGM gas covering a wide
range of ionization states. The 28 star forming galaxies in the main
COS-Halos sample \citep{werk12} range over $9.3<\log\mstar/\msun<10.8$
with a median $\log\mstar/\msun=10.1$; these galaxies lie on the \mzr\
\citep{werk12}.  We therefore consider the COS-Halos results to
constrain the metal budget of CGM gas within 150\,kpc of galaxies at
this typical stellar mass.
Though, as we show, there is little dependence of the observed column
densities (and thus our inferred CGM metal masses) with stellar mass,
the fixed physical radius of 150\,kpc corresponds to $\sim
35$--$110$\% of the virial radius $\rvir$ within this stellar mass
range (\citealt{moster10}, though see \citealt{shull14}).

\begin{figure}
\includegraphics[width=0.48\textwidth]{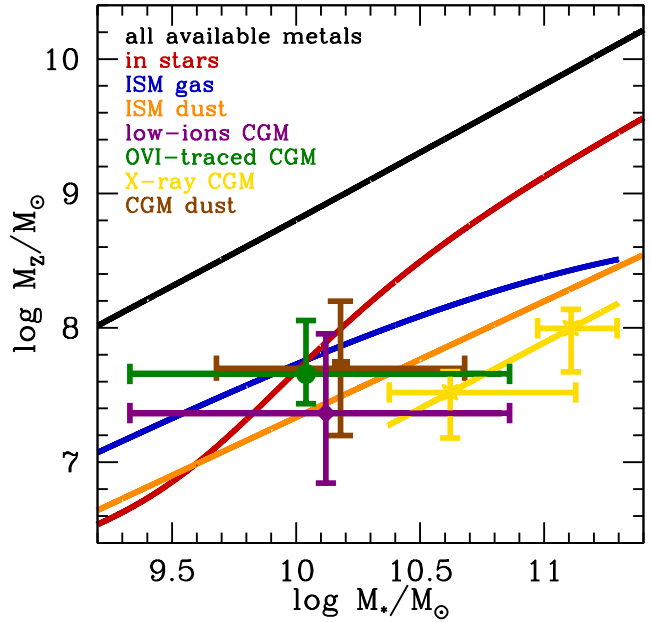}
\caption{\label{fig:mcgm}Masses of metals in the CGM compared to
  galaxy components, with low-ionization CGM shown in purple, the
  \ovi-traced CGM shown in green, CGM dust shown in brown, and the
  X-ray traced CGM shown in yellow. }
\end{figure}

As we show in Figure~\ref{fig:mcgm}, the two predominant phases of the
CGM are a patchy, cool ($\log T \lesssim 5$) low-ionization state gas
(\S\,\ref{sec:low}) and a more highly ionized and uniformly
distributed \ovi-traced gas (\S\,\ref{sec:ovi}).  The much hotter
X-ray traced gas (\S\,\ref{sec:xray}) represents a relatively small
contribution to the overall metal budget of star-forming galaxies.
There is, however, a significant mass of metals in circumgalactic dust
(\S\,\ref{sec:igdust}). Table~\ref{tbl:masses} shows the fiducial,
minimum, and maximum masses at the median COS-Halos galaxy stellar
mass of $10^{10.1}\msun$.
 
\begin{table}
\centering
\begin{tabular}{lrrr}\hline\hline
Component & Fiducial & Minimum & Maximum \\\hline\hline
\multicolumn{4}{c}{Source Components}\\\hline
Type~II SNe & $6.8\phantom{0}$ & $4.9\phantom{0}\phantom{0}$ & $10.5\phantom{0}\phantom{0}$ \\
Type~Ia SNe & $0.76$ & $0.18\phantom{0}$ & $2.9\phantom{0}\phantom{0}$ \\
AGB stars & $0.45$ & $0.22\phantom{0}$ & $0.89\phantom{0}$ \\\hline\hline
\multicolumn{4}{c}{Galactic Components}\\\hline
Stars & $0.72$ & $0.62\phantom{0}$ & $0.79\phantom{0}$\\
ISM gas & $0.81$ & $0.48\phantom{0}$ & $1.1\phantom{0}\phantom{0}$\\
ISM dust & $0.26$ & $0.16\phantom{0}$ & $0.40\phantom{0}$\\\hline\hline
\multicolumn{4}{c}{Circumgalactic Components}\\\hline
Low-ions CGM & $0.23$ & $0.069$ & $0.89\phantom{0}$\\
\ovi-traced CGM & $0.46$ & $0.28\phantom{0}$ & $1.1\phantom{0}\phantom{0}$\\
CGM dust & $0.50$ & $0.16\phantom{0}$ & $1.16\phantom{0}$\\
\end{tabular}
\caption{\label{tbl:masses} 
  Fiducial, minimum, and maximum metal masses for metals made available
  from supernovae and AGB stars, and in each of the galactic and CGM
  components we consider here, for a galaxy with a present-day stellar
  mass of $10^{10.1}\msun$ (the median stellar mass of galaxies in the
  COS-Halos sample).  All units are in $10^{8}\,\msun$. We do not quote
  an X-ray
  traced CGM mass here as this stellar mass is below that of the
  \citet{anderson13} galaxies.
}
\end{table}

\subsection{The Low-Ionization Circumgalactic Medium}\label{sec:low}
\begin{figure}
\includegraphics[width=0.48\textwidth]{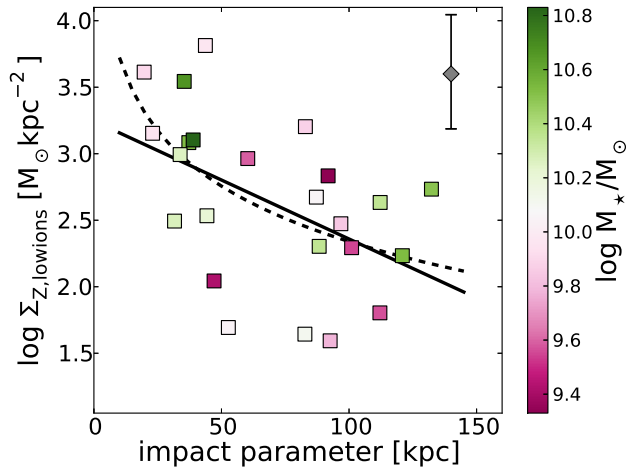}
\caption{\label{fig:lowions}The surface density of low-ionization
  metals as a function of impact parameter and stellar mass; we find
  no evidence of a strong dependence of $\Sigma_{\rm Z}$ on stellar mass. The grey
  point in the upper-right corner shows the typical uncertainty in the
  low-ionization metal surface density. The solid and dashed
  lines
  are fits to $\Sigma_{\rm Z}(\rperp)$ given in
  Equation~(\ref{eqn:slowions}) and (\ref{eqn:plowions}),
  respectively. These fits are subsequently used in
  Equation~(\ref{eqn:mlowions}) to calculate the low-ionization CGM
  metal mass. }
\end{figure}

The mass of metals in the circumgalactic medium in a low-ionization state is
\begin{equation}\label{eqn:mlowions}
M_{\rm Z, lowions} = \int 2\pi\rperp\Sigma_{\rm Z,lowions}(\rperp)\,{\rm d}\rperp,
\end{equation}
where the mass surface density of metals $\Sigma_{\rm Z}$ is 
\begin{equation}\label{eqn:sigma}
\Sigma_{\rm Z, lowions} = ({\rm Si/Z})_{\odot}\times m_{\rm Si}N_{\rm Si},
\end{equation}
where (Si/Z)$_{\odot}=0.04553$ is the Solar mass fraction of metals in
silicon (assuming $12+
\log({\rm Si/H})=7.53$, \citealp{lodders09}), $m_{\rm Si}=28.0855m_p$ is the mass of a silicon atom, and $N_{\rm Si}$
is the ionization-corrected column density of silicon along the line
of sight. We integrate Equation~(\ref{eqn:mlowions}) over the impact
parameter range observed by COS-Halos, from 10 to 150\,kpc.
We summarize here how the silicon column density $N_{\rm Si}$ is derived, with further detail
in Werk et al.\ (in preparation).

For each absorber, we vary the ionization parameter ($\log U$) and the
metallicity to search for {\sc Cloudy} (v08.00, last
described by \citealt{ferland13}) models that are consistent with the
column densities of the \hi\ and metal ions determined from the
observed UV spectra. We assume the gas is photo-ionized, and thus with
temperatures of $\sim 10^{4}$\,K, which is consistent with the Doppler broadening of the line profiles. 
The
ionization parameter is less susceptible to uncertainties in the \hi\
column density than the gas metallicity, making our measurement of the
total metal column density not very sensitive to saturation in the
\hi\ lines, which is common in the COS-Halos sample \citep{tumlinson13}.
  This is mostly because we can independently constrain
$\log U$ from the {\sc Cloudy} models based on several different ionization
states of the same element and the detection of a number of different
metal lines of various ionization states (e.g., \cii, \ciii, and \civ;
\siii, \siiii, and \siiv; \nii\ and \niii; \mgi\ and \mgii).  The ``intermediate ions'', such as \civ\ and \siiv, kinematically trace the
low-ions, and their column densities are generally well-described by the {\sc
  Cloudy} models.  In
obtaining the total metal column along an individual sightline, we
simply apply an ionization correction to the observed ion column
density (e.g., $\log N_{\rm Si III}$ to $\log N_{\rm Si}$) based on
our {\sc Cloudy} analysis; in the few cases lacking robust measurements of
silicon lines, we instead use carbon as a proxy for all metals. We then assume relative Solar abundance ratios to derive
the total metal surface density (Equation~\ref{eqn:sigma}). In cases for which we have observed more than two
different ionization states for a single element, and more than one
element, the uncertainty in the ionization correction along the
individual sightline is low, $< 0.2$\,dex. In cases for which we have
only a single metal ion, or multiple metals of similar ionization
states, the uncertainty in $\log U$ can be over one dex. 
The typical uncertainty in the ionization correction leads to an
uncertainty in the low-ionization metal surface density of
$\pm 0.42$\,dex, as shown in the upper-right corner of Figure~\ref{fig:lowions}.
In general, our derived temperatures, metallicities, ionization
corrections, and subsequently inferred gas masses, are consistent with
those found in similar studies using ionization diagnostics to infer CGM physical conditions. \citep[e.g.,][]{stocke13}.

Figure~\ref{fig:lowions} shows $\Sigma_{\rm Z, lowions}$
as a function of impact parameter and stellar mass\footnote{We have
  converted the \citeauthor{werk12}\ stellar masses to a
  \citeauthor{chabrier03a} IMF from a \citeauthor{salpeter55} IMF.} for the star-forming galaxies in the
COS-Halos sample.  The surface density profile is well-described by
\begin{equation}\label{eqn:slowions}
\log \frac{\Sigma_{\rm Z,lowions}}{\msun\,{\rm kpc}^{-2}} = 3.25 -0.0089\left(\frac{\rperp}{\rm kpc}\right) 
\end{equation}
and
\begin{equation}\label{eqn:plowions}
\log \frac{\Sigma_{\rm Z,lowions}}{\msun\,{\rm kpc}^{-2}} = 5.11 \left(\frac{\rperp}{\rm kpc}\right)^{-1.38};
\end{equation}
the differences in the CGM metal mass derived from either of these fits is negligible.
With the current data, there is no measurably significant dependence
of surface density on the stellar mass of the central galaxy. We,
therefore, derive a low-ionization CGM metal mass of
\begin{equation}\label{eqn:lowions}
\log M_{\rm Z,lowions}/\msun = 7.36 \pm 0.14^{\rm (stat)} \, _{-0.38}^{+0.45 \rm (sys)},
\end{equation}
where the statistical uncertainty is derived by bootstrapping over the
data sample and re-fitting the surface density profile and the
systematic uncertainty owes to the uncertainties in the ionization corrections. 

If other galaxies have high velocity clouds (HVCs;
\citealp{muller63}), such as those seen around the Milky Way, this gas
would be a (small) subset of the low-ionization CGM detected by
COS-Halos.  These discrete gas clouds lie mostly within $\sim 10$\,kpc of the
Galactic plane \citep{wakker07,wakker08,thom08b,hsu11}.  Though their
covering fraction on the sky is high, $\sim$18--67\%
\citep[depending on their \hi\ column density,][]{sembach00,lockman02,shull09,lehner11,shull11b,lehner12,putman12}, their total mass is only $M_{\rm
  hvc}\sim 10^8\,\msun$ (\citealp{shull09,putman12}, though see also
\citealp{lehner11} for a slightly higher estimated mass of ionized
HVCs).  Taking this mass with a fiducial $Z_{\rm hvc}=0.2\zsun$ and
Solar abundance ratios, we find a mass of metals in this
high-velocity circumgalactic component of $M_{\rm Z, hvc}\sim
3\times 10^5\,\msun$, which is small relative to other
circumgalactic (or galactic) components of $L^*$ galaxies.

\subsection{The High-Ionization Circumgalactic Medium}\label{sec:ovi}
\begin{figure*}
\includegraphics[width=0.48\textwidth]{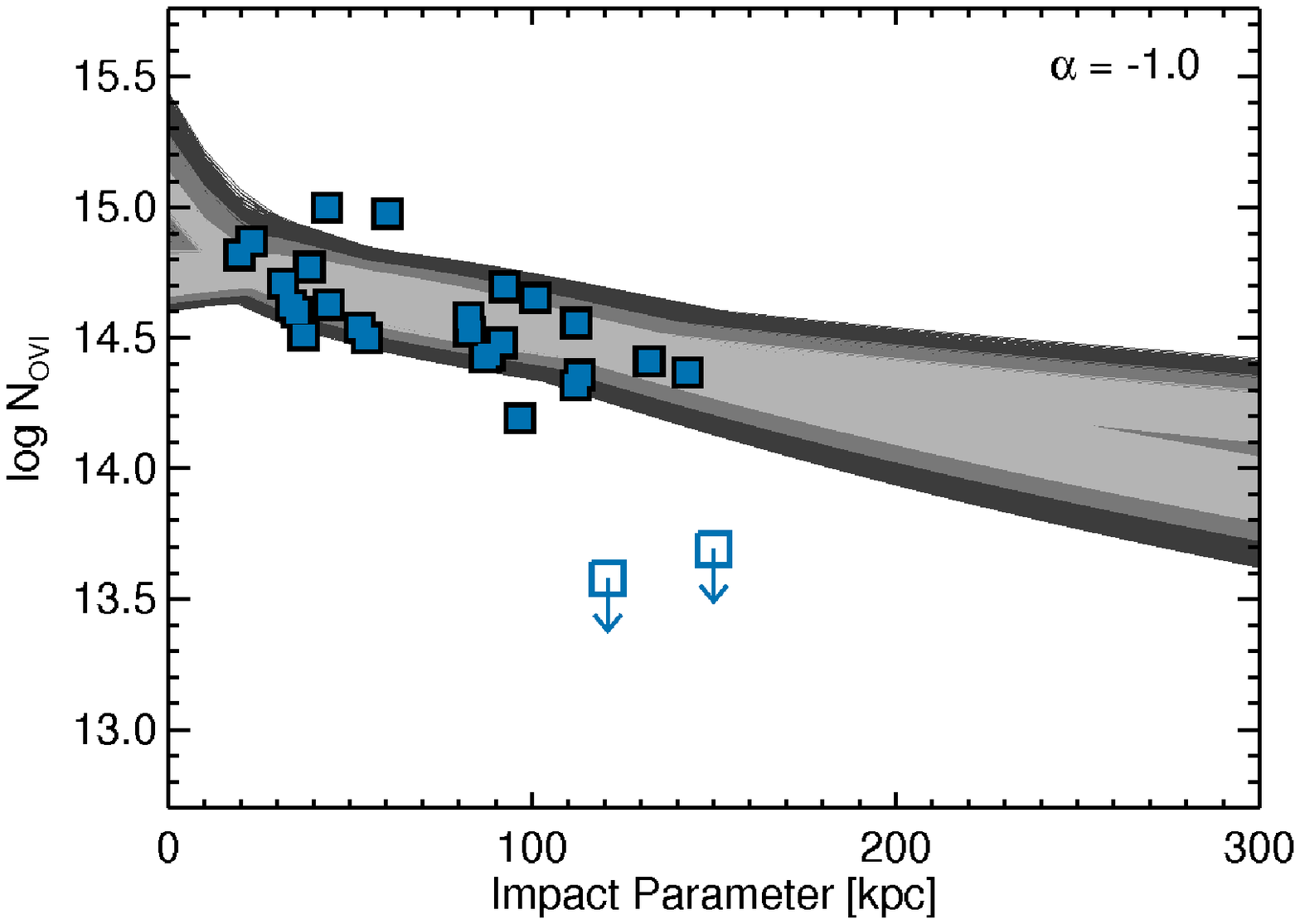}\hfill
\includegraphics[width=0.48\textwidth]{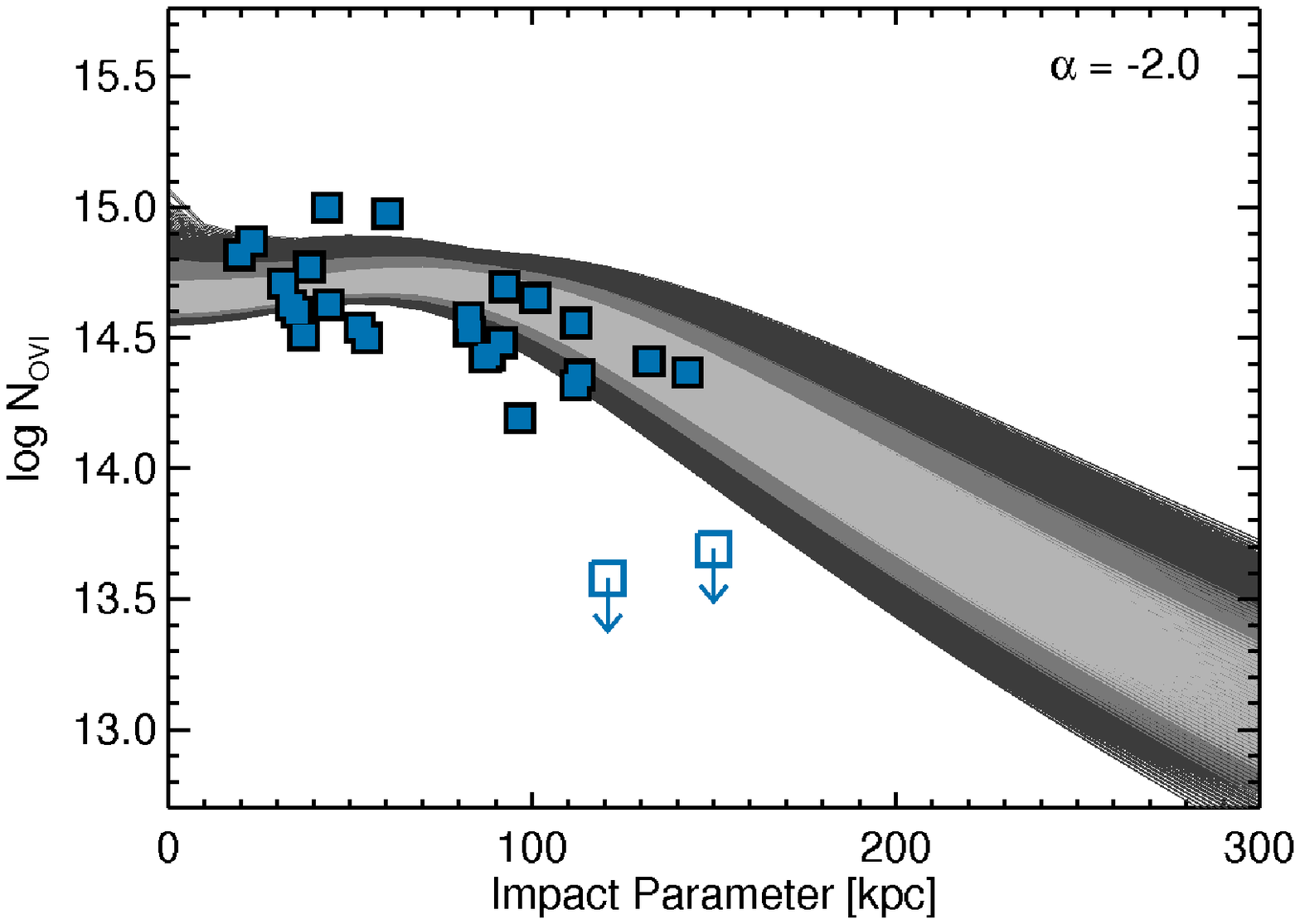}\\
\includegraphics[width=0.48\textwidth]{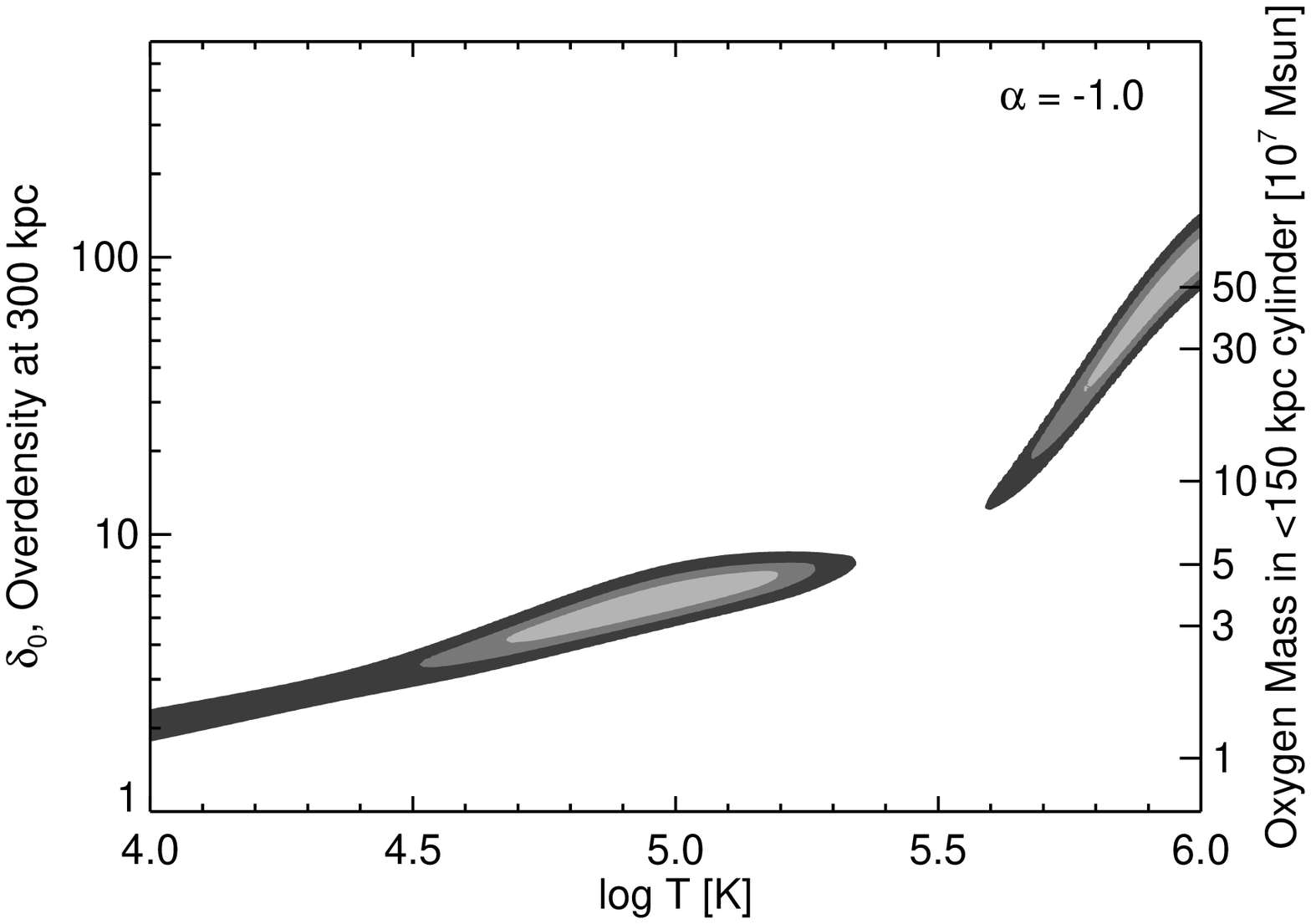}\hfill
\includegraphics[width=0.48\textwidth]{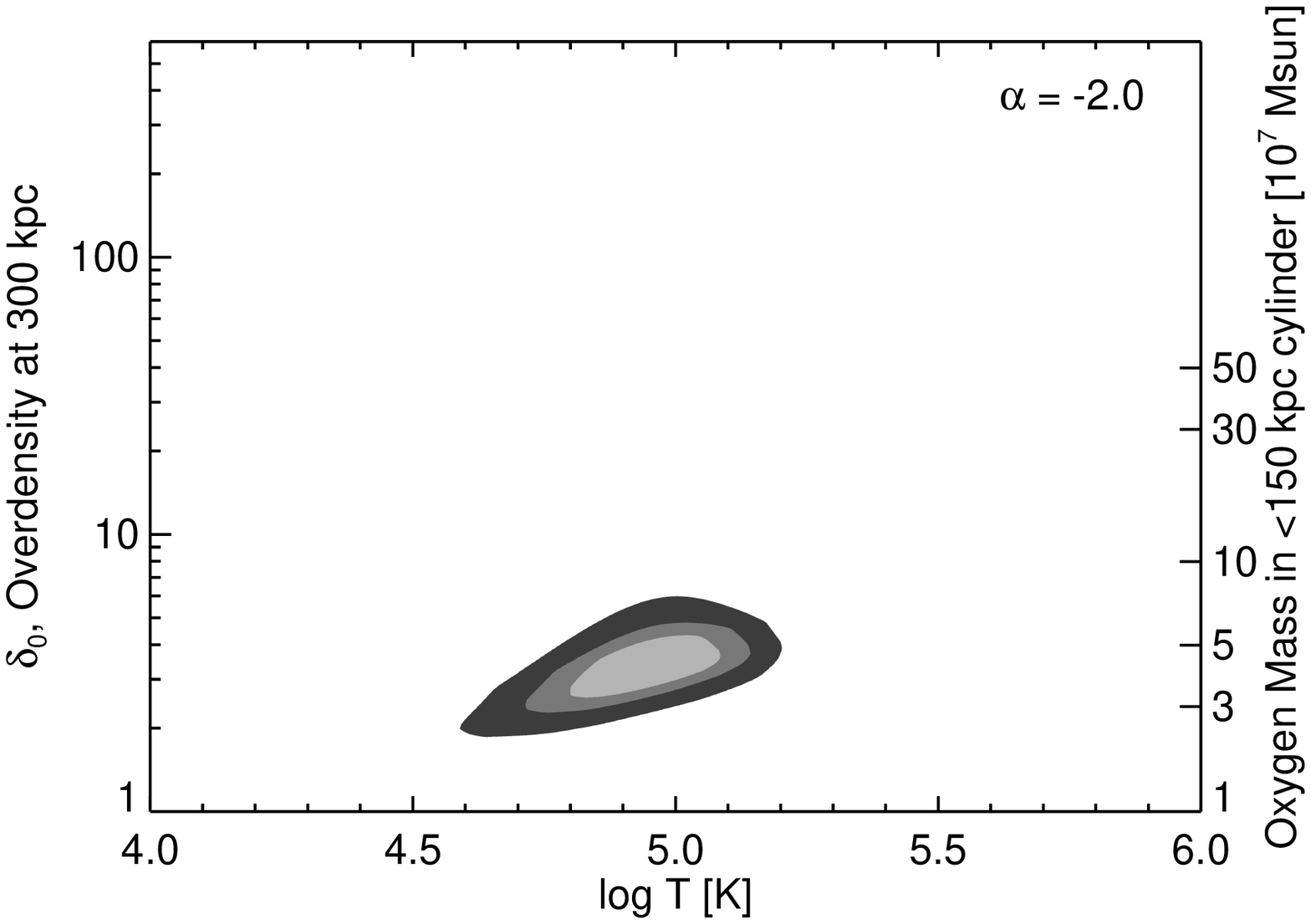}
\caption{\label{fig:halos}Top panels: \ovi\ column density profiles
  from simple halo models showing the range of profiles in agreement
  with the data at 1, 2, and 3 $\sigma$ confidence, with $\alpha=-1$
  (left column) and $-2$ (right column).  Detections of \ovi\ around
  the COS-Halos star-forming galaxies are overlaid in blue.  Higher
  impact parameter data from \citet{prochaska11c} prefer the
  $\alpha=-2$ model.  
  Bottom panels: Corresponding temperatures and overdensities for the
  simple halo models. The
  oxygen masses in cylinders of radius 150\,kpc are degenerate with
  the density normalization.
}
\end{figure*}

\citet{tumlinson11} found $N_{\rm O\,VI}\sim 10^{14.5}$\,cm$^{-2}$
around star-forming galaxies out to impact parameters of 150\,kpc (as
shown in the top panels of Figure~\ref{fig:halos}). For the most part,
this 
\ovi\ absorption is kinematically
distinct from the low-ionization gas, and its high column density is inconsistent with the \ovi-traced gas
being in the same ionization state, temperature, and density as the
low-ionization gas (\S\,\ref{sec:low}; Werk et al.\ in
preparation); we therefore consider the material traced by this \ovi\
absorption as an additional reservoir of metals.

Using a similar method to the one we used in \S\,\ref{sec:low},
\citet{tumlinson11} found an \ovi-traced oxygen mass of $M_{\rm
  oxy,OVI}\ge 1.2\times 10^{7}\,\msun$. This estimate is a direct
numerical integration over column densities and surveyed volumes, and
assumes the maximum \ovi\ ionization fraction $f_{\rm OVI}=0.2$
achieved in photo- and/or collisional ionization to set a lower limit
on the column density and mass of total oxygen.  In this simple
integration, $M_{\rm oxy,OVI}\propto f_{\rm OVI}^{-1}$, so the total
oxygen mass can increase substantially above this conservative minimum
if the detected \ovi\ is not at its optimal conditions for
ionization. In particular, the maximum \ovi\ ionization fraction
$f_{\rm OVI}=0.2$ occurs at $T\sim 10^{5.5}$\,K, where cooling times
are short and therefore such a massive reservoir of gas is presumably short-lived.

To impose a conservative limit on the CGM oxygen mass,
\citet{tumlinson11} did not consider oxygen away from its optimal
ionization fraction in \ovi , and did not consider gas outside the
150\,kpc region of the COS-Halos survey.  To assess how much more
oxygen mass there could be out to $\rvir$ than the conservative
minimum, we consider simple \oned\ halo models wherein we take into
account not only the overall level of \ovi\ absorption, but also its
profile as seen through projection.  In these descriptive models, we
take the density profile of the CGM to be a power-law with a slope
$\alpha$, normalized to an overdensity $\delta_0$ at a radius
$R_0=300$\,kpc, i.e.,
\begin{equation}\label{eqn:halo}
\delta = \delta_0 \left(\frac{R}{300\,{\rm kpc}}\right)^{\alpha},
\end{equation}
with the inner 10\,kpc sphere (which would not be considered part of a
halo) set to zero density to prevent numerical divergence.
The \ovi\ temperature is set to a constant $T$.  This \oned\ model is
then projected into a 3-D spherical distribution.  Mock sightlines are
then passed through this medium at a fixed projected separation
(impact parameter) and column densities of \ovi\ are obtained via line
integrals along the sightline. The models assume that the diffuse
gaseous medium is exposed to the \citet{haardt01} radiation
background, and includes a component for collisional ionization
equilibrium  at the parametric temperature. These models were
also used in \citet{tumlinson13} to assess the total \hi\ masses in
galaxy halos from the COS-Halos \hi\ survey.

The primary parameters in this model are the exponent of the density
power law $\alpha$, the density normalization at $R_0=300$\,kpc,
$\delta_0$, and the gas temperature $T$. To estimate the amount of
\ovi, and from that the oxygen masses in the model halos, we produce
grids of models with $T=10^{4}$--$10^{6}$\,K and $\alpha = -1$ or
$-2$. We then use a simple likelihood analysis to find the set of
parameters that best match the \ovi\ detections for the COS-Halos
star-forming sample. The allowable model profiles for $\alpha = -1$
and $-2$ appear in the top panels of Figure~\ref{fig:halos}; the
bottom panels show the contours of total oxygen mass in the $\delta_0$
v.\ $T$ parameter space. Here we see that the low-temperature ($T\sim
10^{5}$\,K) solution exists for both values of $\alpha$ and gives a
total mass to 150 kpc of 3 to $5 \times 10^7\msun$, a few times higher
than the conservative minimum.  

The main controlling parameter is the power-law slope of the density
profile, for which we have no independent constraint; if we take the
much shallower profile of $\alpha=-1$, then a much more massive
($M_{\rm oxy,OVI}\sim 2\times 10^8\msun$), higher density
($\delta_0\sim 100$), hotter ($T\sim 10^6$\,K) solution becomes
possible. The choice of $R_0$ affects the best-fit overdensity, but
does not have a large impact on the implied mass within 150\,kpc.  For the mass of
oxygen traced by \ovi\ within a cylinder of radius 150\,kpc, we
therefore take
\begin{equation}\label{eqn:ovi}
M_{\rm oxy,OVI} = \left\{  
\begin{array}{ll}
1.2\times 10^{7}\,\msun, & \mbox{minimum,} \\
2\times 10^{7}\,\msun, & \mbox{fiducial, and} \\
5\times 10^{7}\,\msun, & \mbox{maximum.} \\
\end{array}\right.
\end{equation}
To derive a metal mass, we assume a Solar oxygen-to-metals ratio of
44\%, as done for the ISM in \S\,\ref{sec:zism}.  As with the
low-ionization metals, we note that
we do not see a dependence of $N_{\rm OVI}$ on stellar mass within the
1.5\,dex mass range in $\mstar$ spanned by our sample.

It is possible that the \ovi-traced gas is in a transient phase,
potentially cooling out of a hotter reservoir or ``boiling off'' of
cooler clouds; either way, one would expect for the reservoir the
\ovi-traced gas is transitioning out of to be more massive than the
\ovi-traced gas itself. This simple picture, however, is at odds with
the relatively lower masses of other CGM components, unless, e.g., the
``hotter'' component is $< 5.8\times 10^6$K (i.e., 0.5\,keV) and
therefore not readily detected in the X-rays. By
design, our ``\ovi-traced'' mass does not address such a ``hidden''
reservoir, which would only add to the CGM metal mass budget. 

\subsection{The X-Ray Traced Circumgalactic Medium}\label{sec:xray}
\citet{anderson13} used stacked ROSAT images to place constraints on
the X-ray luminosity and thus hot ($T\gtrsim 6\times 10^6$\,K) CGM gas mass
out to $\sim 50$\,kpc.  They find that late-type galaxies with $L_K =
1.35^{+0.72}_{-0.36}\times 10^{11}\lsun$ have a hot X-ray halo with
gas mass $M_{\rm hot gas}=3.6^{+0.8}_{-1.1}\times 10^{9}\msun$ for a
metallicity of $0.3\zsun$; as their systematic uncertainty in the gas
mass scales roughly linearly with the assumed metallicity, we adopt
$0.3\zsun$.  For fainter late-type galaxies, with $L_K =
4.4^{+9.7}_{-1.9}\times 10^{10}\lsun$, extended X-ray emission is not
robustly detected in the stacked images, although they give an
estimate of the hot CGM mass of $M_{\rm hot
  gas}=1.2^{+0.5}_{-0.6}\times 10^{9}\msun$.  Furthermore,
\citeauthor{anderson13}\ argue that there cannot be a {\em massive}
reservoir of hot gas at galactocentric radii of $\gtrsim
50$\,kpc. These hot CGM masses are consistent with other probes of hot
gaseous halos around $\sim L^{*}$ galaxies, including the dispersion
measure toward pulsars in the Large Magellanic Cloud, the pressure
confinement of HVCs in the Milky Way's halo, the (lack of) \ion{O}{7}\
absorption in the halos of other galaxies, and X-ray surface
brightness limits from individual galaxies (\citealp{anderson10}, though
see also \citealt{fang13}).  Likewise, \citet{yao10} did not detect X-ray absorption from a
plethora of metal lines in stacked Chandra
grating spectra in 12 CGM sightlines, strongly suggesting that hot
$\zsun$ gas cannot fill a typical CGM volume.
We stress that this picture---in which
there is a relatively small amount of material in a coronal X-ray
traced phase---is dramatically different from what is expected for
galaxies living in more massive halos with higher virial temperatures,
especially rich galaxy clusters.

Following \citet{anderson13}, in Figure~\ref{fig:mcgm}, we show the
mass of metals in a hot phase extrapolated out to 150\,kpc around late-type
galaxies to be 
\begin{equation}\label{eqn:xray}
\log M_{\rm hotcgm,\,Z}/\msun = 0.98(\log\mstarnot/\msun - 11) + 7.89,
\end{equation}
adopting a $K$-band mass-to-light ratio of
$\mstar/L_K=0.95\,\msun/\lsun$ \citep{bell03}. This represents the
fiducial \citet{anderson13} values increased by a factor of 6 to
account for out to 150\,kpc. This extrapolation is uncertain as it
depends on the unknown slope of the hot halo profile
\citep{kaufmann09,feldmann13}; we, therefore, let this mass ratio
extrapolation factor vary from 3 to 8.\footnote{We thank M.\ Anderson
  for calculating the extent of this extrapolation to 150\,kpc.} Finally, we note that the
stellar mass range of the galaxies in the \citet{anderson13} sample is
$\sim 1$\,dex higher than those in the COS-Halos sample.

\subsection{Dust outside of galaxies}\label{sec:igdust}
Measuring the systematic reddening of background quasars relative to
their projected distance from foreground galaxies from SDSS,
\citet{menard10} derived a CGM dust mass of $M^{\rm cgm}_{\rm
  dust}\simeq 5\times 10^7\msun$ for $20h^{-1}{\rm kpc}<r_{\rm
  eff}<r_{\rm vir}$ around $\sim 0.5L^*$ galaxies.  This mass is
consistent with the dust expected to be associated with the level of
\mgii\ absorption seen in the COS-Halos sample
\citep{menard08,menard12,werk13}.  
Though this dust mass is calculated for out to $\rvir$, as opposed to
the 150\,kpc we adopt for the other CGM reservoirs,  more recent results
are showing that CGM dust is primarily confined to within 150\,kpc of
galaxies (Peek \etal, in preparation).
We adopt $\log M^{\rm cgm}_{\rm
  dust}/\msun = 7.7\pm 0.5$ for a stellar mass range of $9.7 \leq
\log\mstar/\msun \leq 10.7$. 

The \citeauthor{menard10} average galaxy redshift is $\simeq 0.36$, so
by $z=0$ these dust masses could grow somewhat larger. However, this
typical redshift is only slightly above that of the COS-Halos median
redshift.  Finally, we note that this $M^{\rm cgm}_{\rm dust}$
combined with the COS-Halos results (\S\S\ref{sec:low}--\ref{sec:ovi})
implies that $\sim 42$\% of CGM metals by mass are in a solid phase,
which is remarkably close to the simulation results of \citet{zu11}, who
found that a dust mass fraction of $0.39$ was required to to reproduce
\citet{menard10} results in a momentum-driven wind model that also
reproduces observed galaxy properties and intergalactic enrichment.
(Intriguingly, this dust mass ratio of 0.42 is slightly higher than the interstellar
ratio of $\sim 0.3$ in the same $\mstar$ range.)

\section{How bad is the missing metals problem?}\label{sec:missing}
We address here the so-called ``missing metals problem'': does the sum
total of metals found in the reservoirs explored here, i.e., those
inside galaxies (\S\,\ref{sec:gals}) and those in the observed
circumgalactic medium (\S\,\ref{sec:cgm}), fully account for the
available budget of metals produced (\S\,\ref{sec:yields})?  The
answer to this question largely depends on the quantity of metals that galaxies
have produced. As shown in Figure~\ref{fig:flost}, the uncertainties
in the nucleosynthetic yields are large enough that they can easily
dominate this question: we must know how many metals we are looking
for before we can determine whether or not they have all been found.
For the discussion here we will consider as the available metal budget
just the fiducial values discussed in \S\,\ref{sec:yields}, with the
caveat that more accurate yields could easily change our conclusions
one way or the other.

\begin{figure*}
\includegraphics[width=0.48\textwidth]{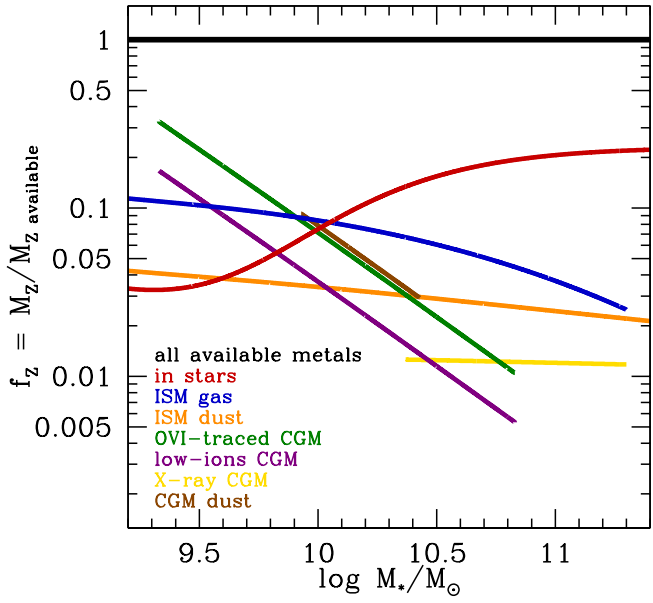}\hfill
\includegraphics[width=0.48\textwidth]{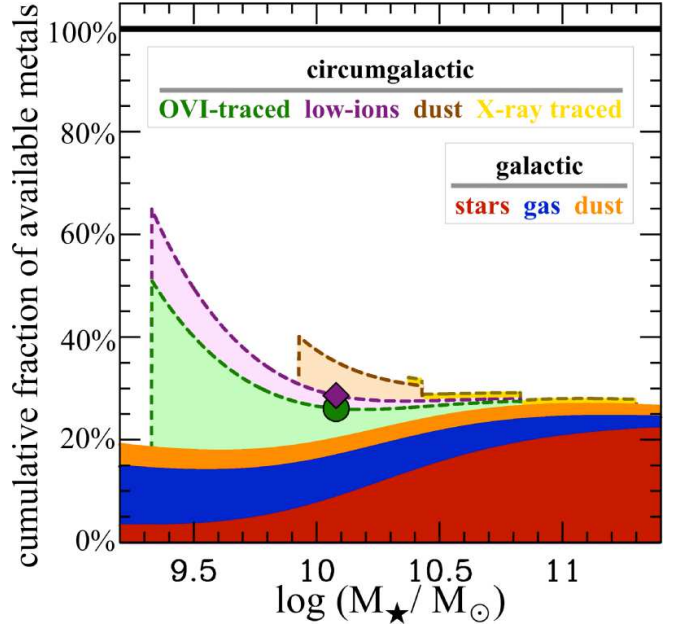}
\caption{\label{fig:frac}Fractions ({\em left}) and cumulative
  fractions ({\em right}) of available metals in various components as
  a function of stellar mass. Note that the CGM masses are within a
  fixed {\em physical} radius of 150\,kpc---and thus sampling different fractions
  of the virial radius with respect to stellar mass.}
\end{figure*}

In Figure~\ref{fig:frac}, we show the fraction of metals that were
ever produced and subsequently expelled
by supernovae and stellar winds as currently observed in each of the
components in each of the three galactic (stars, interstellar gas and
dust; \S\,\ref{sec:gals}) and the four major CGM (low-ionization,
\ovi-traced, dust, X-ray traced, each out to 150\,kpc; \S\,\ref{sec:cgm}) components we
consider.  That is, we produce the left panel of Figure~\ref{fig:frac}
by dividing the fiducial values in Figure~\ref{fig:mcgm} by
the fiducial ``available'' metals line.  The right panel plots the
same information, but as cumulative fractions.  Only a few percent of
the available
metals are in interstellar dust, $\sim 5$--$10$\% are in interstellar
gas, and, in the galactic stellar mass range that we consider, the massive galaxies
have more metals trapped in stars than the smaller galaxies have in
the ISM.  It is also immediately obvious from Figure~\ref{fig:frac}
that at $\log\mstar/\msun \lesssim 10$ the observed CGM could
dominate the metal budget, though at higher masses it is still
comparable to the metals found in galaxies.  The sharp fractional
increase in the \ovi-traced and low-ionization CGM stems from the fact
that we are dividing constant CGM masses by a steeply changing mass of
available metals; we caution that though our data does not indicate a
dependence of CGM metal mass within 150\,kpc on stellar mass, this is
both a fixed {\em physical} radius, and we are limited by relatively
small numbers.

\begin{figure*}
\includegraphics[width=0.48\textwidth]{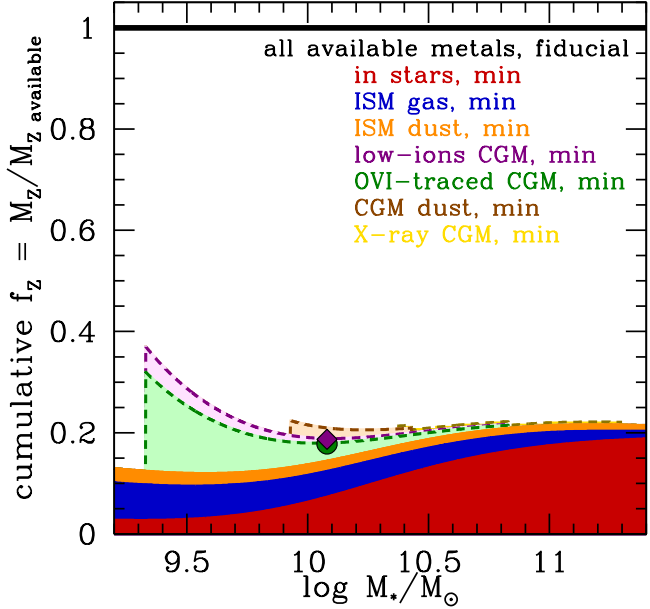}\hfill
\includegraphics[width=0.48\textwidth]{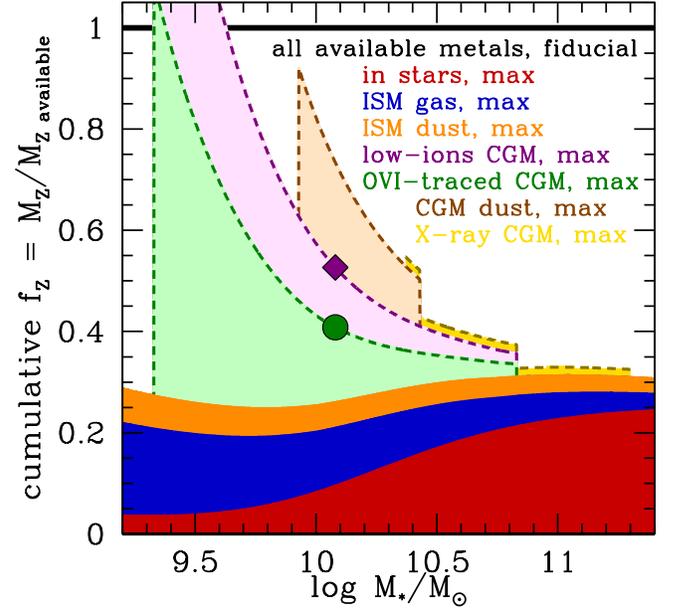}
\caption{\label{fig:pessopt}Pessimistic ({\em left}) and optimistic
  ({\em right}) cumulative metal fractions. In the ``pessimistic''
  case, we set the values of the mass of metals in each galactic and
  circumgalactic component to the minimum value allowed within our
  uncertainties; in the ``optimistic'' case, we take the maximum
  allowed values. In both cases, we compare to the fiducial values for available metals.}
\end{figure*}

In the fiducial case, we find that at about a Milky Way mass, roughly
40\% of all available metals are easily accounted for when including
measurements of the CGM out to $150$\,kpc.  Figure~\ref{fig:pessopt}
shows the ``pessimistic'' and ``optimistic'' cases for all metals,
i.e., where we set the masses of each of the individual components to
either their minima or maxima, respectively, while keeping the
available budget at its fiducial value.  Here, we see that at the
median mass of the COS-Halos sample, $\log\mstar/\msun\sim 10.1$, the
uncertainties imply that we do not account for between 25 to 70
percent of metals.  Within galaxies, the largest source of this
uncertainty is from the uncertainty in the gas-phase metallicities,
especially at low $\mstar$.

Reducing the uncertainty in the interstellar abundance calibrations
would help constrain the fraction of metals remaining in galaxies; it
is possible that new emission line diagnostics adopting a
$\kappa$-distribution for the electron energies could resolve some of
the systematic uncertainties currently plaguing this field, leading to
smaller uncertainties in the near future \citep{nicholls12,dopita13}.
Resolved \hii-region metallicity mapping, such as the upcoming Mapping
Nearby Galaxies at APO (MaNGA) program with SDSS-IV, combined with
resolved gas maps could also reduce the need to make strong
assumptions about how well the ISM is mixed.

With regards to the UV-traced CGM, until we can resolve this gas in
emission and accurately map its morphology and extent for individual
galaxy halos, absorption line studies are the only way to detect and
characterize this massive reservoir. Given the apparent patchiness of
the low-ionization CGM, with only $\sim 40$ sightlines, the COS-Halos
sample is not large enough to detect trends in CGM properties with
respect to galaxy mass. In particular, it is unclear with the
absorption line approach how much of the large scatter in column
density (Figure~\ref{fig:lowions}) is galaxy-to-galaxy variation as opposed to
genuine CGM substructure.
Fortunately, the metallic CGM is
easier to quantify than the baryonic CGM;
as most of the observable
transitions are from metal ions probing a large temperature range, no
metallicity correction is needed, and a large hidden reservoir of
predominantly ionized 
metal-poor gas will affect the metal census less than the baryonic
one.

For the other CGM components, our inventory relies on a statistical
measure of the CGM dust mass, but it is possible for the same kind
of reddening studies to target different masses of foreground
galaxies (Peek \etal, in preparation).  Likewise, the \citet{menard10}\ analysis does not constrain
the chemical makeup of the circumgalactic dust (see further discussion
in the Appendix).  Finally, we note that we currently must assume a
metallicity of the X-ray traced CGM component instead of directly
measuring a metal mass, but as this component is relatively small
around the galaxies that we are considering, the contribution of this
uncertainty to the total metal census uncertainty is not large. A
larger contributing factor is the uncertainty in the slope of the
X-ray profile, especially beyond 150\,kpc, as a very flat profile
could lead to a total hot halo mass that is several times larger than
we quote here \citep{kaufmann09,crain13,feldmann13,anderson13,fang13}.

As metals (e.g., \ovi) are observed at impact parameters beyond
150\,kpc \citep{prochaska11c, stocke13}, we do not expect that even in
the most optimistic scenario for our current accounting to have found
all of the metals. For example, an extrapolation of the \ovi\ column
densities following the $\alpha=-2$ profile in Figure~\ref{fig:halos}
to an impact parameter of 300\,kpc is consistent with the \ovi\
observations from \citet{prochaska11c}.  Despite an increase in volume
of a factor of eight, extrapolating this profile from 150 to 300\,kpc would
only roughly double the (spherical) \ovi-traced CGM metal mass.

\section{Conclusions and Discussion}\label{sec:conc}
We have compiled an accounting of metals in and around star-forming
$\sim L^*$ galaxies at $z\sim 0$ and compared this census to the
available budget of heavy elements produced and expelled by supernovae
and stellar winds. We find:
\begin{enumerate}
\item Galaxies with stellar masses $10^9$--$10^{11.5}\msun$ retain in
  their stars, ISM, and dust a roughly constant $\sim 20$--$25$\% of
  the metals they have produced. Thus, the bulk of metals outside of galaxies
  at $z=0$ were produced by (the precursors to) today's $\sim L^*$
  galaxies.
\item About half of the metals produced by typical star-forming
  galaxies can be accounted for by considering the CGM out to $\sim
  150$\,kpc when added to the stellar and interstellar components. Most of these circumgalactic metals are in a highly
  ionized \ovi-traced phase or in dust.
  The low-ionization CGM metal mass is subdominant,
  though not negligible. We find the metals in a circumgalactic
  hot gas component are small compared to ISM contributions, based on
  the hot gas halo mass estimates of \citet{anderson13} with a
  metallicity of $0.3\zsun$.
\item With the current data, there is no evidence for a steep dependence on
  galaxy mass in the metal surface density in the CGM (within this
  stellar mass range). Though the metal surface
  density does decline with impact parameter, the
  sightline-to-sightline variation for the low-ionization CGM is large,
  suggesting that the low-ionization CGM is patchy \citep[cf.][]{werk13}.
\item The largest source of uncertainty in the ``missing metals
  problem'' is from the uncertainties in the nucleosynthetic yields;
  to know what fraction of metals we have found, we must first know
  how many metals we are seeking.
\end{enumerate}

There are two important improvements that should be made to this
$z\sim 0$ metal accounting: finding the ``missing'' metals, and
expanding the census 
to other galaxy populations.  Within galaxies,
unaccounted-for metals could be in a warm, ionized component of the
ISM (i.e., the WIM, \S\,\ref{sec:ism}). In the CGM, metals could be
hiding in a hotter reservoir from which the \ovi-traced gas might be cooling,
provided it is still too cool to be detected in X-ray emission (i.e.,
$T\lesssim$1--5$\times 10^6$\,K). Constraints on such a reservoir could be
placed by observing the CGM in higher ionization transitions such as
\neviii\ or \mgx\ \citep{savage11,narayanan11,ford13}. Finally, as discussed in \S\,\ref{sec:missing}, our
current inventory is by construction missing the metals at impact
parameters $> 150$\,kpc \citep{prochaska11c}. While the UV-traced
CGM could be mapped using QSO absorption lines as done in the
COS-Halos survey, a next-generation X-ray telescope such as the
International X-ray Observatory (IXO; \citealp{bookbinder12}) would be
required to detect, e.g., low-column density \ovii\ expected to be
found at high impact parameters \citep{anderson13,hummels13}.

The two main galaxy populations lacking from our inventory are
passive galaxies and dwarf galaxies.  \citet{gallazzi08} found that as
much as $\sim 40$\% of the metals produced by bulge-dominated galaxies
could be currently residing in their stars.  The circumgalactic metal
content of passive galaxies with $\log\mstar/\msun\sim 10.5$ to $11$
can begin to be addressed with COS-Halos.  Intriguingly, while the
properties of the low-ionization state absorbers around the passive
galaxies are statistically similar to those around star-forming
galaxies \citep{thom12,werk13}, passive galaxies have
extremely little \ovi\ in their halos \citep{tumlinson11}.  More
massive galaxies than those probed by the COS-Halos survey are also
expected to reside in halos with higher virial temperatures, and so it
is also reasonable to expect that more massive passive galaxies 
have a larger fraction of their circumgalactic gas traced by X-ray
emitting rather than UV-absorbing gas. On the other hand, the
star-formation histories and IMFs (and IMF histories), and thus
available metal budgets, are somewhat more difficult to constrain for
passive galaxies.  Likewise, while their dust content should be
negligible, passive galaxies do have an interstellar medium.  This
ISM, though, is generally not cold, and---by definition---not lit up
by \hii\ regions, making its metal content difficult to ascertain.  It
is possible, though, that passive galaxies with low-ionization
emission (i.e., {\sc Liners}) could have their interstellar metal
content probed with analogous methods \citep{yan12}.

Dwarf galaxies pose a somewhat different and interesting set of
obstacles to understanding the eventual fate of their metals.  From
the results we show here, it is obvious that the uncertainties in the
gas masses will be much more relevant for $\mstar<10^{9.5}\,\msun$
galaxies, and, simultaneously, it is less clear to what extent
the ISM of dwarf galaxies is mixed, and thus whether or not
equation~(\ref{eqn:ism}) is valid.  Moreover, in star forming
galaxies, it appears that the dust in dwarf galaxies is higher than
the metal mass locked up in stars; as this conclusion is based on very
few galaxies, a full census of metals in
star-forming dwarf galaxies will require a better understanding of
their dust content.  Moreover, dwarf galaxies have had more stochastic
star formation histories \citep{weisz11} and appear to have larger
spread in gas-phase metallicities \citep{zahid12a} than $\sim L^*$
galaxies.  Some of this stochasticity might be attributable to
environment, as metallicities \citep{pasquali10,pasquali12} and star
formation (i.e., passive versus star-forming; \citealp{geha12})
properties strongly depend on the degree of isolation.  With
regards to the CGM of dwarf galaxies, our group has another large {\em
  HST} program, COS-Dwarfs (PID 12248, 129 HST orbits, PI:\ J.\
Tumlinson), designed to map the CGM of $8\lesssim
\log\mstar/\msun\lesssim 9.5$ galaxies out to $\sim 150$\,kpc in \hi,
\cii, \civ, \sii, \siii\, \siiv, and other species, with which we will
address these questions.


\acknowledgements{We thank M.\ Shull for a thoughtful, detailed, and rapid referee's report.
  We are grateful to M.\ Anderson, K.\ Barger, E.\ Bell, R.\ Bordoloi,
  M.\ Childress, M.\ Dopita, A.\ B.\ Ford, A.\ Fox, E.\ Jenkins, J.\
  C.\ Howk, J.\ Kalirai, N.\ Lehner, S.\ Mathur, B.\ M\'enard, S.\ Oey, J.\ E.\
  G.\ Peek, R.\ S.\ Somerville, C.\ Thom, and S.\ Trager for useful
  and interesting conversations that have improved this paper.
  Support for program GO11598 was provided by NASA through a grant
  from the Space Telescope Science Institute, which is operated by the
  Association of Universities for Research in Astronomy, Inc., under
  NASA contract NAS 5-26555.
  MSP acknowledges support from the Southern California Center for
  Galaxy Evolution, a multi-campus research program funded by the
  University of California Office of Research. This research has made
  extensive use of NASA's Astrophysics Data System.  }

\appendix
\section{An accounting of Oxygen}
As the production sites of the different elements
(Figure~\ref{fig:elements}) differ, the eventual fate of
different elements should also differ
\citep{tinsley79,matteucci86}. We attempt here to address how the
census of the most abundant heavy element, oxygen, differs from that of
``all metals''. In most cases, we have either derived metal masses
from oxygen (e.g., ISM gas or the \ovi-traced CGM), and / or we lack
the sensitivity to non-Solar abundance patterns (e.g., the
low-ionization or X-ray traced CGM). With both stars
(\S\,\ref{sec:oxystar}) and dust (\S\,\ref{sec:dustoxy}), however, we
can begin to address the elemental distribution of metals.

\subsection{Oxygen in Stars}\label{sec:oxystar}
Where the stellar abundances of $\alpha$-elements (of which oxygen is
one, along with, e.g., Mg, Si, Ne, etc.) have been measured relative
to stellar Fe abundances, massive galaxies have been found to be more
$\alpha$-enhanced than less massive galaxies \citep[e.g.,][]{worthey92,thomas05}.  As
done by \citet{peeples13} we therefore renormalize the
\citet{gallazzi05} data according to the relation fit to data from \citet{arrigoni10},
\begin{equation}\label{eqn:alphafe}
\alphafe = 0.085 + 0.062(\log\mstar/\msun - 10),
\end{equation}
with a $\sigma$-to-$\mstar$ conversion as measured by
\citealt{thomas05}, where $\alphafe$ is the logarithm of the ratio of
the abundance of $\alpha$-elements to the abundance of iron, where
$\alphafe=0$ is the solar ratio. We note that the normalization of
this relation has decreased as the reference models have improved
\citep{arrigoni10,conroy13b}.

We apply this correction by assuming that the metals-to-iron and
$\alpha$-to-oxygen ratios follow Solar abundance patterns. 

We adopt a solar oxygen abundance of
$[12+\log(\mbox{O}/\mbox{H})]_{\odot}=8.76$ \citep{caffau11}.  The
uncertainty plotted in right-hand panel of Figure~\ref{fig:mz}
includes the range of older estimates of 8.69 \citep{asplund09} to
8.89 \citep{delahaye06}.  We further include the range in uncertainty for
letting $\alphafe=0$ for all galaxies; for most of this mass range,
this is similar to adopting the higher oxygen abundance with a varying
$\alpha/{\mbox{Fe}}$ ratio.  Because this correction is larger for
more massive galaxies, the inferred uncertainty in the oxygen mass in
stars is also larger.

\subsection{Oxygen in Dust}\label{sec:dustoxy}
We assume that 27\%\ of dust is oxygen, by mass, with a possible error
range of 22--35\%.
Using the \citet{weingartner01}, \citet{zubko04}, and
\citet{draine09} dust models to constrain the oxygen fraction of dust
\citep[see also][]{draine11}, we find
\begin{equation}\label{eqn:dustfrac}
\frac{\moxydust}{\mdust} = \frac{\mu_{\rm oxy} N^{\rm oxy}_{\rm dust}}{\sum\limits_X^{} \mu_{X} N^X_{\rm dust}},
\end{equation}
where $\mu$ is the average atomic mass, $X$ is one of the elements C,
N, O, Mg, Si, and Fe, and $N$ denotes the number abundance.  In these
three models, $\moxydust/\mdust$ is 25.1, 32.1, and 25.7\%,
respectively.
We note that
observational constraints on the oxygen mass fraction of dust from
oxygen dust depletions are difficult to reconcile with chemical models
\citep{sofia01,draine03,jenkins09,whittet10}. Moreover, depletion studies generally imply that
the oxygen-to-metals ratio in dust is as high as in the Sun.  

As there are many other uncertainties we do not explicitly take into
account, we show in Figure~\ref{fig:mz} a range of 22--35\% with a
best-guess of 27\%, after adopting the mean dust relation given in
Equation~(\ref{eqn:dust}).  

Finally, we note that this dust oxygen fraction of 27\% is less than
the Solar oxygen-to-metals ratio of 44\%.

\subsection{Oxygen lost from galaxies}
\begin{figure*}
\includegraphics[width=0.48\textwidth]{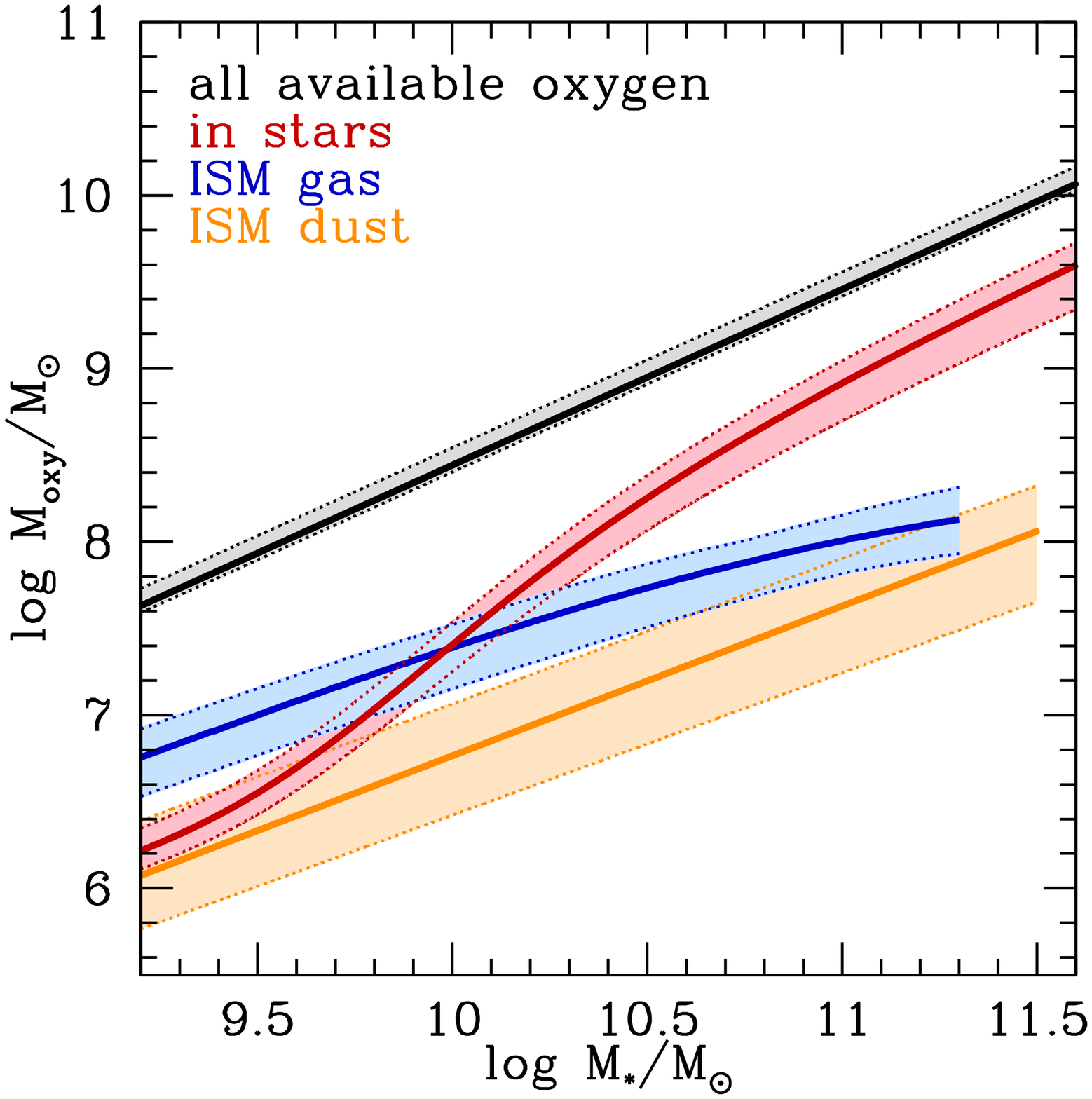}\hfill
\includegraphics[width=0.48\textwidth]{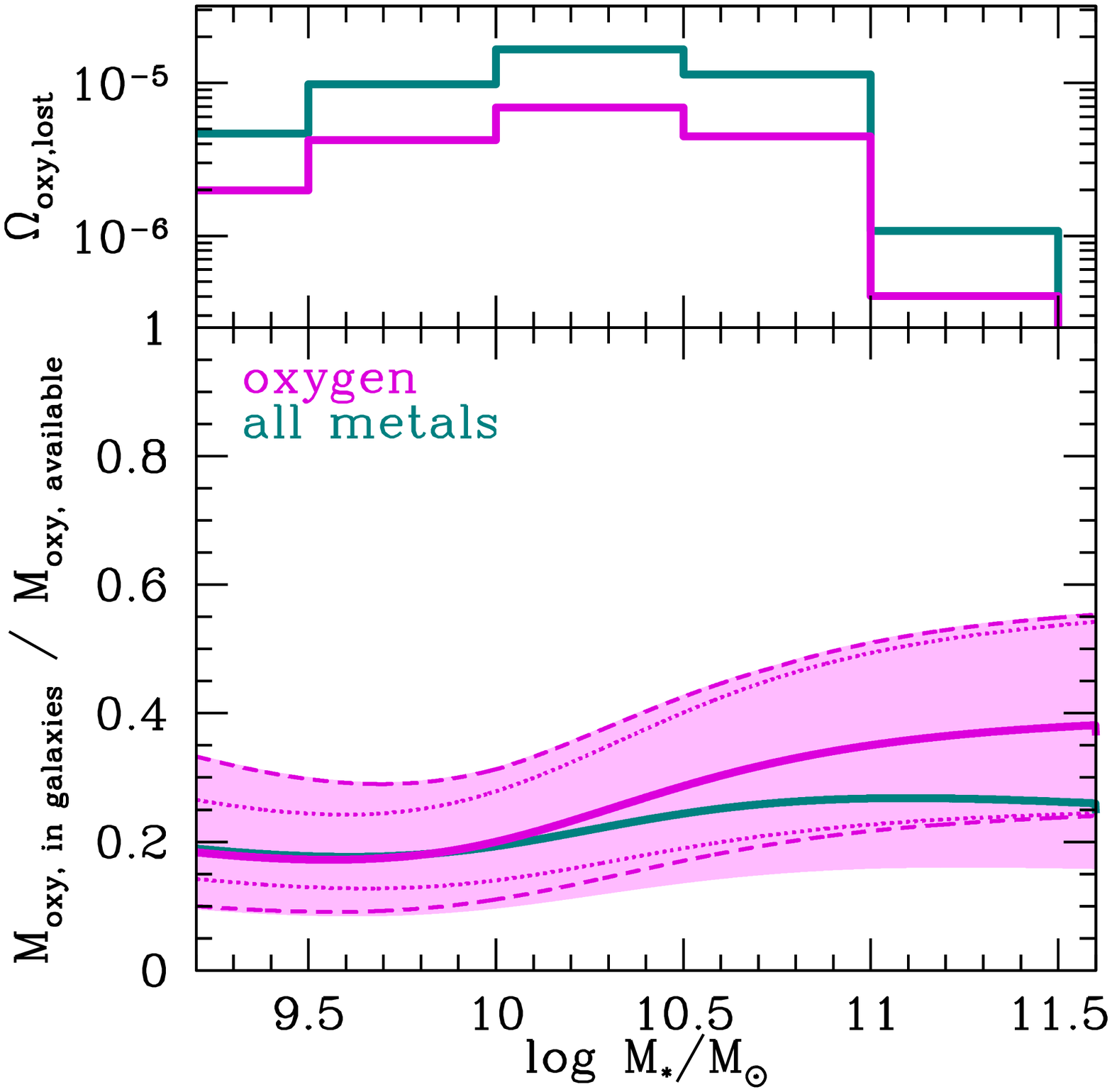}
\caption{\label{fig:moxy}{\em Left:} Same as Figure~\ref{fig:mz}, but
  for just oxygen. {\em Right:} Same as Figure~\ref{fig:flost}, but
  for oxygen. The increase in
  $f_{\rm oxy,retain}$ at high $\mstar$ is nearly entirely due to the
  increase in $\alphafe$ in stars.}
\end{figure*}

We consider the oxygen lost from galaxies,
\begin{equation}\label{eqn:oxylost}
\Omega_{\rm oxy,\,lost}  = \frac{1}{\rho_{\rm c}}\int\limits^{M_{\star,{\rm max}}}_{M_{\star,{\rm min}}}M_{\rm oxy,\,lost}(\mstar)\;\frac{{\rm d}\,n(\mstar)}{{\rm d}\,\log\mstar}\,{\rm d}\,\log\mstar,
\end{equation}
where
\begin{equation}\label{eqn:moxylost}
M_{\rm oxy,\,retain}(\mstar) = \mstaroxy + \moxyism + \moxydust.\nonumber
\end{equation}
The right panel of Figure~\ref{fig:moxy} shows $\foxyretain\equiv M_{\rm
  oxy,\,retain}/M_{\rm oxy,\,made}$ v.\ $\log\mstar$. The increase in
$\foxyretain$ relative to the metals available fraction is entirely
driven by the increase in the stellar $\alphafe$ with stellar mass;
unlike for metals, the uncertainties in this fraction are not
dominated by the yield uncertainties, but rather roughly equal
contributions of stellar and ISM, with the uncertainty in the oxygen
dust fraction entering at larger stellar masses.  This large
difference is particularly surprising since if the abundance ratios of
star-formation driven outflows differ from solar, they are probably
$\alpha$-enhanced, implying that $\alphafe$ {\em outside} of galaxies
should also be greater than in the sun.  One possible solution to this
conundrum is that the ISM might not have solar abundance ratios, or
have $\alphafe$ ratios that also depend on stellar mass; here, we
measure the ISM {\em oxygen} mass, but infer its metal mass by
assuming a solar oxygen-to-metals ratio.  If the ISM is
$\alpha$-depleted for the $\mstar\gtrsim 10^{10.5}\msun$ galaxies,
then the shape of $\fzretain(\mstar)$ might have a shape similar to
$\foxyretain(\mstar)$ rather than being so surprisingly flat.

\subsection{Circumgalactic Oxygen and the ``Missing Oxygen Problem''}
\begin{figure*}
\includegraphics[width=0.48\textwidth]{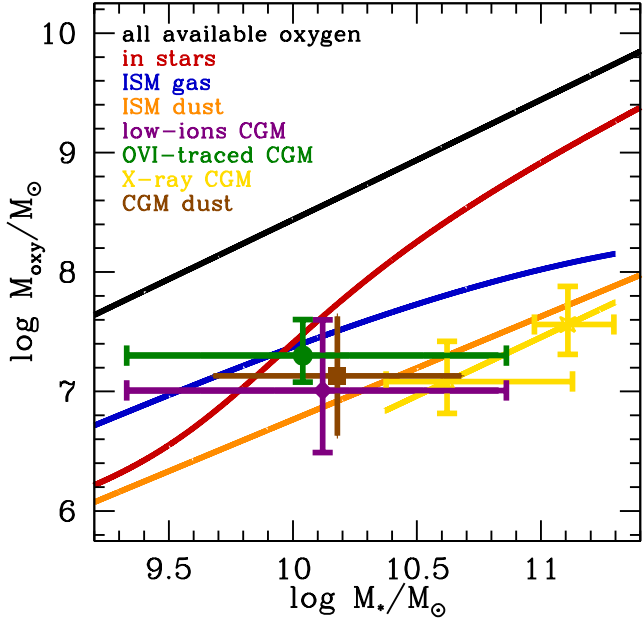}\hfill
\includegraphics[width=0.48\textwidth]{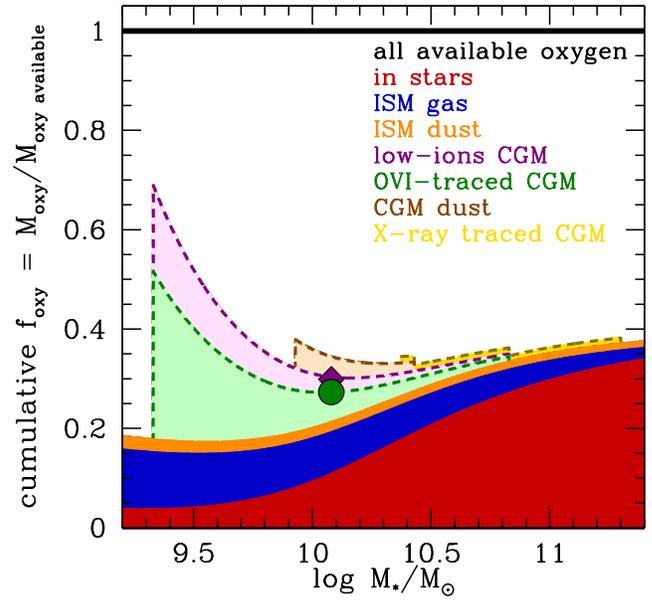}
\caption{\label{fig:oxyfrac} Same as Figure~\ref{fig:frac} but
  for oxygen.}
\end{figure*}

If outflows are preferentially $\alpha$-enhanced (e.g., because they
are driven by Type~II supernovae), then the CGM might have a higher
Oxygen-to-metals ratio than found in the Sun.  The $\alphafe$ ratio
for circumgalactic gas, however, is not yet able to be measured, and
so for the time being we must assume Solar Oxygen-to-metals ratios. An
``$\alpha$-enhanced'' CGM would decrease our \ovi-derived metal
masses, but it is superficially unclear how a higher SN~II
contribution would affect the metal masses we derive for the
low-ionization CGM.  As shown in Figure~\ref{fig:oxyfrac}, we take the
\ovi-traced oxygen mass given in Equation~(\ref{eqn:ovi}), and take a
Solar oxygen-to-metals ratio of 0.44 for the low-ionization and X-ray
traced CGM components (though see \citealt{origlia04}).  For
circumgalactic dust, we assume CGM dust has a similar chemistry to
that of ISM dust, with a dust oxygen mass fraction of 27\%
(\S\,\ref{sec:dustoxy}). (Given the uncertainties in how dust can
survive in first a wind fluid and subsequently in the CGM, it is
unlikely that circumgalactic dust has a similar chemistry to
interstellar dust, but there is currently no conclusive evidence one
way or the other.)

\begin{figure*}
\includegraphics[width=0.48\textwidth]{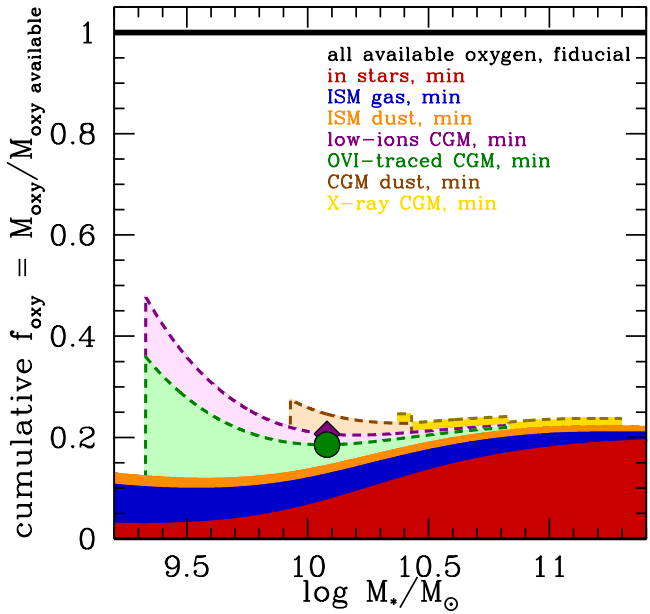}\hfill
\includegraphics[width=0.48\textwidth]{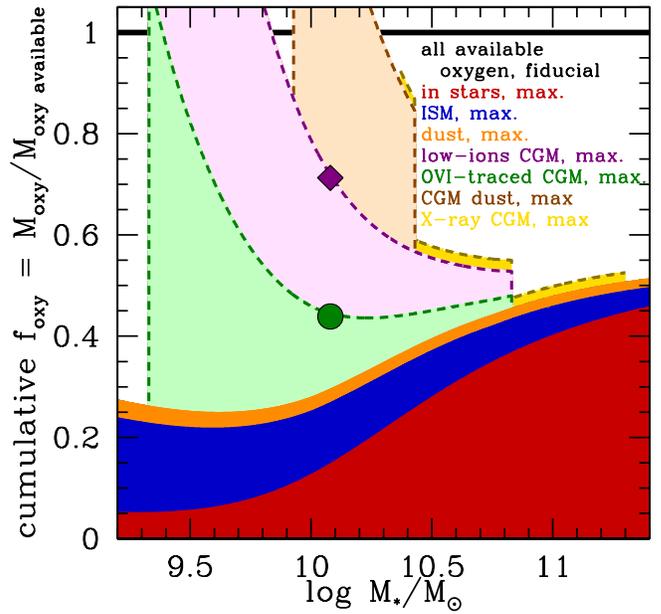}
\caption{\label{fig:oxypessopt}Same as Figure~\ref{fig:pessopt} but
  for oxygen.}
\end{figure*}

With regards to the ``missing oxygen problem'' (Figure~\ref{fig:oxypessopt}),
the picture is a little more complicated than for all metals.
The uncertainties in the oxygen masses at the
high-$\mstar$ end are dominated by the uncertainty in the
stellar $\alphafe$ in star-forming galaxies.  The observed dependence
on the stellar $\alpha$-enhancement in passive galaxies
\citep{thomas05,arrigoni10} is generally attributed to an age dependence.  Within
the mass range of the star-forming population we consider here,
however, there is still expected to be an age-mass relation that
should also manifest as a correlation between $\alphafe$ and
$\mstar$, though it would be surprising if this relation is the
same as for passive galaxies.  Likewise, if supernovae are significant
drivers of redistributing metals into the CGM, then it is quite
possible that the CGM could be $\alpha$-enriched while the ISM is left
$\alpha$-depleted; the current observations are not sensitive to these
differences.  

\end{document}